\documentclass[journal]{IEEEtran}
\usepackage{amsmath, amsthm, amsfonts, amssymb}
\usepackage{algorithmic}
\usepackage{algorithm}
\usepackage{array}
\usepackage{xcolor}
\usepackage[caption=false,font=footnotesize,labelfont=rm,textfont=rm]{subfig}
\usepackage{textcomp}
\usepackage{stfloats}
\usepackage{url}
\usepackage{verbatim}
\usepackage{makecell}
\usepackage{graphicx}
\usepackage{cite}
\usepackage{bm}
\usepackage{booktabs}
\usepackage[T1]{fontenc}
\usepackage{xspace}
\usepackage[utf8]{inputenc}

\DeclareMathAlphabet\mathbfcal{OMS}{cmsy}{b}{n}

\begin{document}

\title{Age-Gain-Dependent Random Access for Event-Driven Periodic Updating}

\author{Yuqing Zhu, Yiwen Zhu, Aoyu Gong, Yan Lin, Yuan-Hsuan Lo, and Yijin Zhang
\thanks{This work was supported in part by the National Natural Science Foundation of China under Grant 62071236. 
(\emph{Corresponding author: Yijin Zhang}.)}
\thanks{Yuqing Zhu, Yiwen Zhu, Yan Lin, and Yijin Zhang are with the School of Electronic and Optical Engineering, Nanjing University of Science and Technology, Nanjing 210094, China (e-mail: \{yuqing.zhu; zyw; yanlin\}@njust.edu.cn; yijin.zhang@gmail.com).}
\thanks{Aoyu Gong is with the School of Computer and Communication Sciences, \'Ecole Polytechnique F\'ed\'erale de Lausanne, Lausanne 1015, Switzerland (e-mail: aoyu.gong@epfl.ch).}
\thanks{Yuan-Hsuan Lo is with the Department of Applied Mathematics, National Pingtung University, Pingtung 90003, Taiwan (e-mail: yhlo@mail.nptu.edu.tw).}
}

\maketitle

\begin{abstract}
This paper considers utilizing the knowledge of age gains to reduce the average \textit{age of information} (AoI) in random access with event-driven periodic updating for the first time. Built on the form of slotted ALOHA, we require each device to determine its age gain threshold and transmission probability in an easily implementable decentralized manner, so that the contention can be limited to devices with age gains as high as possible. For the basic case that each device utilizes its knowledge of age gain of only itself, we provide an analytical modeling by a multi-layer \textit{discrete-time Markov chains} (DTMCs), where an external DTMC manages the jumps between the beginnings of frames and an internal DTMC manages the evolution during an arbitrary frame, for obtaining optimal access parameters offline. For the enhanced case that each device utilizes its knowledge of age gains of all the devices, we require each device to adjust its access parameters for maximizing the estimated network expected AoI reduction per slot, through maintaining a posteriori joint probability distribution of local age and age gain of an arbitrary device in a Bayesian manner. Numerical results validate our study and demonstrate the advantage of the proposed schemes over other schemes.

\end{abstract}

\begin{IEEEkeywords}
Internet of Things, age of information, periodic update, random access, slotted ALOHA.
\end{IEEEkeywords}

\section{Introduction}
\subsection{Background}
Internet of Things (IoT) systems have been widely applied in many real-time services~\cite{Mehdi2018,Zheng2019,Luvisotto2019}, such as emergency surveillance, target tracking, process control, and so on.
In these services, destinations are interested in the status of one or multiple processes observed by multiple sources, and then take necessary actions based on the received status updates.
To ensure the quality and even safety of these services, it is typically necessary for sources to deliver their generated updates to the corresponding destinations as timely as possible.

However, such timeliness requirement \emph{cannot} be characterized adequately by conventional performance metrics (e.g. throughput and delay).
For example, when the throughput is large, the received updates may not be fresh due to long delay; when the delay is small, the received updates may not be fresh due to infrequent arrivals of updates. 
As such, a new performance metric, termed \textit{age of information} (AoI), has been introduced in~\cite{2011Minimizing} to measure the time elapsed since the generation moment of the latest successfully received update at a destination.
Naturally, to reduce the\textit{ network average AoI} (AAoI), it is desirable for multiple access schemes to assign higher transmission priorities to devices with higher age gains, where the age gain of a device in a slot quantifies how much a successful transmission of this device will reduce its corresponding instantaneous AoI.
Note that the age gain of a device depends on only its instantaneous AoI under the \textit{generate-at-will} (GAW) arrival of updates.

With this objective, scheduling schemes that operate in a centralized manner without contentions have been designed to perform close to optimal network AAoI in various scenarios~\cite{Gong2020Globecom, ACM2020, Maatouk2021TWC, Kadota2021ToM}. 
However, they may be impractical to implement due to the huge overhead of required coordination, especially when there is considerable uncertainty on the arrival patterns of updates.
Unlike scheduling schemes, random access schemes (e.g. slotted ALOHA, frame slotted ALOHA, CSMA) allow a population of devices with limited or no coordination to dynamically and opportunistically share a channel.
So, it is strongly required to design \textit{age-gain-dependent random access} (AGDRA) schemes, where each device utilizes its knowledge of age gains to determine when to transmit its updates in an easily implementable decentralized manner, so that the unavoided contention can be limited to devices with age gains as high as possible.

Various AGDRA schemes have been proposed for the GAW arrival~\cite{Atabay2020INFOCOM,Yavascan2021JSAC,chen2020age,Ahmetoglu2022,Yavascan2023,Zhu2023,Yang2023,Xie2023} and the Bernoulli arrival~\cite{Sun2020TCOM,Chen2022TIT,Moradian2024TCOM,Poorya2024TMC} of updates, and have been shown to significantly reduce the network AAoI compared to conventional random access schemes.
It can be observed from~\cite{Xie2023,Sun2020TCOM,Atabay2020INFOCOM,Yavascan2021JSAC,chen2020age,Ahmetoglu2022,Yavascan2023,Zhu2023,Yang2023,Poorya2024TMC,Chen2022TIT,Moradian2024TCOM} that designing AGDRA is uniquely challenging due to the inherent coupling of the arrival process of updates, the time evolution of local ages, the time evolution of AoIs, and the time-varying mutual interference.
Generally speaking, this coupling would become more complicated when a more general arrival process of updates is considered, and is quite different from that for optimizing the throughput or delay metric.

\subsection{Related Work}\label{related work}
Without relying on the knowledge of age gains, many conventional random access schemes have been proposed for minimizing the network AAoI.
Under the GAW arrival,~\cite{Yates2017ISIT} showed that using slotted ALOHA is worse than scheduling by a factor of about $2e$.  
Under the Bernoulli arrival,~\cite{Kadota2021} used the elementary renewal theorem to optimize the transmission probabilities for slotted ALOHA and CSMA, while~\cite{Wang2023} used \textit{discrete-time Markov chains} (DTMCs) to optimize the frame length for frame slotted ALOHA.
Under the periodic arrival,~\cite{bae2022age} analyzed the effect of maximizing the instantaneous throughput on the network AAoI of slotted ALOHA.

Basic AGDRA, where each device utilizes its knowledge of age gain of only itself to adjust its access parameters, has been investigated in~\cite{Sun2020TCOM,Poorya2024TMC,Atabay2020INFOCOM,Yavascan2021JSAC,chen2020age,Zhu2023,Yang2023,Ahmetoglu2022,Yavascan2023}.
In the form of slotted ALOHA,~\cite{Atabay2020INFOCOM,Yavascan2021JSAC,chen2020age,Zhu2023,Yang2023,Ahmetoglu2022,Yavascan2023} assumed that each device adopts a fixed transmission probability if its corresponding age gain reaches a fixed threshold, but keeps silent otherwise.
Based on a comprehensive steady-state analysis of the DTMC defined in~\cite{Atabay2020INFOCOM}, closed-form expressions of the network AAoI and optimal access parameters were provided in~\cite{Yavascan2021JSAC} for an infinitely large network size.
For an arbitrary network size,~\cite{chen2020age} analyzed the network AAoI by modeling the AoI evolution of each device as a DTMC, which, however, relies on an ideal assumption that the states of all the devices are independent of each other. 
To mitigate the negative impact of contentions, a reservation phase ahead of actual data transmission is proposed in~\cite{Ahmetoglu2022,Yavascan2023}, but its benefit comes at a cost of additional overhead compared to~\cite{Atabay2020INFOCOM,chen2020age,Yavascan2021JSAC}.
Further,~\cite{Yang2023,Zhu2023} used stochastic geometry tools to derive the network AAoI under the spatiotemporal interference.
Note that~\cite{Atabay2020INFOCOM,chen2020age,Yavascan2021JSAC,Zhu2023,Yang2023,Ahmetoglu2022,Yavascan2023} mainly focused on the GAW traffic, except that~\cite{Yavascan2021JSAC} extended its findings to obtain an upper bound on the network AAoI for the Bernoulli arrival. 
These schemes~\cite{Atabay2020INFOCOM,Yavascan2021JSAC,chen2020age,Zhu2023,Yang2023,Ahmetoglu2022,Yavascan2023} have a common advantage that the access parameters can be obtained offline through analytical modeling, and thus can be simply implemented.
In addition, heuristic methods proposed in~\cite{Sun2020TCOM,Poorya2024TMC} allow each device to use different transmission probabilities for different cases, but lack analytical modeling.

To further reduce the network AAoI, enhanced AGDRA, where each device utilizes its knowledge of age gains of all the devices to adjust its access parameters, has been investigated in~\cite{Chen2022TIT,Moradian2024TCOM} for the Bernoulli arrival.
In the form of slotted ALOHA, the AAT proposed in~\cite{Chen2022TIT} allows each device to transmit with a dynamic transmission probability (determined by the estimated number of active devices) for maximizing the instantaneous network throughput only if its corresponding age gain reaches a threshold, which could be computed adaptively using the estimated distribution of age gains.
In the form of frame slotted ALOHA, the practical T-DFSA proposed in~\cite{Moradian2024TCOM} allows the frame length and age-gain threshold for each frame to be adjusted by the estimated distribution of age gains, so that the estimated expected number of active devices can be the smallest number not smaller than a certain number (searched by simulations).
Different from~\cite{Chen2022TIT,Moradian2024TCOM}, under the GAW arrival,~\cite{Xie2023} estimated the network AoI rather than the individual age gains for heuristically adjusting the transmission probability, which would obviously lead to the AoI degradation.

However, these previous studies on AGDRA~\cite{Sun2020TCOM,Atabay2020INFOCOM,Yavascan2021JSAC,chen2020age,Ahmetoglu2022,Yavascan2023,Zhu2023,Yang2023,Chen2022TIT,Moradian2024TCOM,Xie2023,Poorya2024TMC} have not considered the event-driven periodic arrival of updates, which usually appears in many monitoring services\cite{Fu2018TAC,Claudia2011,Lei2021,Feng2019}.
For example, in closed-loop process control, multiple sensors are employed to measure the plant outputs and validate the event conditions periodically, and then each sensor sends a fresh status update to a machine controller as needed.
Note that such an arrival process can include those considered in\cite{Sun2020TCOM,Atabay2020INFOCOM,Yavascan2021JSAC,chen2020age,Ahmetoglu2022,Yavascan2023,Zhu2023,Yang2023,Chen2022TIT,Moradian2024TCOM,Poorya2024TMC,bae2022age,Yates2017ISIT,Kadota2021,Wang2023,Xie2023} as particular cases.

\subsection{Contributions}

To fill the gap in this field, this paper attempts to design a type of AGDRA in the form of slotted ALOHA with an age gain threshold~\cite{Atabay2020INFOCOM,Yavascan2021JSAC,chen2020age,Zhu2023,Yang2023,Ahmetoglu2022,Yavascan2023,Chen2022TIT,Moradian2024TCOM}, called T-AGDSA, under event-driven periodic updating. 
Compared to the existing work~\cite{bae2022age,chen2020age,Yavascan2021JSAC,Chen2022TIT,Moradian2024TCOM}, this paper makes the following key contributions. 
\begin{enumerate}
\item \textit{Basic T-AGDSA}: For simple implementation, consider fixed threshold and fixed transmission probability as in~\cite{Atabay2020INFOCOM,Yavascan2021JSAC,chen2020age,Zhu2023,Yang2023,Ahmetoglu2022,Yavascan2023}. 
We provide an analytical modeling approach to evaluate the network AAoI under event-driven periodic updating, based on which optimal threshold and optimal transmission probability can be obtained offline. 
Compared to~\cite{bae2022age,chen2020age,Yavascan2021JSAC}, the technical difficulty of our work is to consider the mutual impact of event-driven periodic updating and the age-gain-dependent behavior in modeling, which is overcome by a multi-layer DTMC model. Here an external DTMC manages the jumps between the beginnings of frames, while an internal DTMC manages the evolution during an arbitrary frame. 
Note that the modeling approaches in~\cite{bae2022age,chen2020age} can be seen as special cases here.
\item \textit{Enhanced T-AGDSA}: To pursue lower network AAoI, we propose an enhanced T-AGDSA scheme that allows each device to adjust the threshold and the transmission probability for maximizing the estimated network \textit{expected AoI reduction} (EAR) per slot, based on the knowledge of age gains of all the devices. 
Such knowledge comes from using  Bayes' rule to keep a posteriori joint probability distribution of local age and age gain of an arbitrary device.
Compared with the AAT~\cite{Chen2022TIT} that maximizes the estimated instantaneous network throughput under a reasonably controlled effective sum arrival rate, our scheme can avoid low efficiency of the throughput-EAR conversion.  
Compared with the practical T-DFSA~\cite{Moradian2024TCOM} that controls the estimated expected number of active devices reasonably, our scheme can avoid the network EAR degradation when the probability distribution of the estimated number is divergent.    
\end{enumerate}
Through extensive numerical experiments, we validate our theoretical analysis and demonstrate the advantage of the proposed schemes over the schemes in~\cite{bae2022age,Yavascan2021JSAC,Chen2022TIT,Moradian2024TCOM} in a wide range of network configurations.

The remainder of this paper is organized as follows. 
The system model, the considered two versions of T-AGDSA, and a lower bound on the network AAoI are specified in Section~\ref{sec:SystemModel}.
In Section~\ref{sec:SAGRA}, we provide an analytical modeling approach to evaluate basic T-AGDSA for determining optimal fixed access parameters.
In Section~\ref{sec: A practical Scheme}, we propose an enhanced T-AGDSA scheme for maximizing the estimated network EAR per slot. 
Section~\ref{Sec:Numerical Results} provides numerical results to verify our study. 
Section~\ref{Sec:Conclusion} draws final conclusions.

\section{System Model and Preliminaries}  \label{sec:SystemModel}

\subsection{Network Model}\label{sec: network model}
\label{sec:NetworkModel}
\begin{figure}[!htbp]
	\centering
	\includegraphics[width=3.4in]{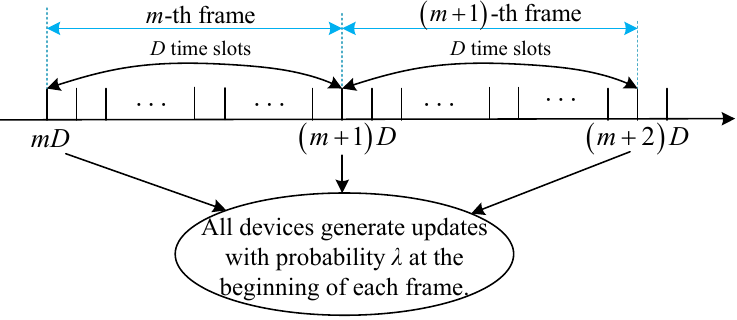}
	\caption{Time Structure.}
	\label{fig: TimeStructure}
\end{figure}

Consider a globally-synchronized uplink IoT system consisting of a common \textit{access point} (AP) and $N$ devices, indexed by $\mathcal{N}\triangleq\{1,2,..., N\}$.
As shown in Fig.~\ref{fig: TimeStructure}, the global channel time is divided into frames (indexed from frame 0), each of which consists of $D$ consecutive slots.
The slots in frame $m$ are indexed from slot $mD$ to $(m+1)D-1$.
At the beginning of each frame, each device independently generates a single-slot update with probability $\lambda\in(0,1]$ and does not generate updates at other time points.
To maintain the information freshness, a newly generated update at each device will replace the undelivered older one if there is any.

By considering a reliable wireless channel under an appropriate modulation and coding scheme, we assume that an update is successfully transmitted if it is not involved in a collision, and otherwise is unsuccessfully transmitted.
After a successful reception of an update of a device, the AP immediately sends an \textit{acknowledgment} (ACK) to notify the device without errors or delays.
Thus, at the end of each slot $t$, each device is able to be aware of the channel status of slot $t$, denoted by $c_t\in \{0 (\text{idle}), 1 (\text{success}), * (\text{collision})\}$.

\subsection{Performance Metrics}\label{sec: Performance Metrics}

At the beginning of slot $t$, we denote the local age of device $n$ by $w_{n,t}$, which measures the number of slots elapsed since the generation moment of its freshest update.
The local age of device $n$ is reset to zero if the device generates a new update at the beginning of slot $t$, otherwise, it increases by one. 
Then, the evolution of $w_{n,t}$ with $w_{n,0} = 0$ is given by
\begin{equation}
	\label{eq:Evolution_w}
	{w_{n,t+1}} =
	\begin{cases}
		{0,} &\text{if device $n$ generates an update}\\ 
             &\text{at the beginning of slot $t+1$,} \\
		{w_{n,t}+1,}&{\text{otherwise.}}
	\end{cases}
\end{equation}

Next, we denote the instantaneous AoI of device $n$ at the beginning of slot $t$ by $h_{n,t}$, which measures the number of slots elapsed since the generation moment of its most recently successfully transmitted update.
If the freshest update of device $n$ is transmitted successfully at slot $t$, the AoI of device $n$ will be set to its local age (in the previous time slot) plus one, otherwise, the AoI will increase by one. 
Then, the evolution of $h_{n,t}$ with $h_{n,0} = 0$ is given by
\begin{equation}
	\label{eq:Evolution_h}
	{h_{n,t+1}} =
	\begin{cases}
		{w_{n,t}+1},&\text{if device $n$ successfully }\\ 
                    &\text{transmits in slot $t$,}\\
		{h_{n,t}+1},&{\text{otherwise.}}
	\end{cases}
\end{equation} 
Owing to the ACK mechanism and local information about $w_{n,t}$, each device $n$ is able to be aware of the value of $h_{n,t+1}$ at the beginning of slot $t+1$ for each $t \ge 0$.

We deﬁne the AAoI of device $n$ as:
\begin{equation}
\label{deqn_ex2a}
    {\Delta}_n\triangleq\lim_{T\to\infty}\frac{1}{T}\sum_{t=0}^{T-1}h_{n,t}. 
\end{equation}
This paper aims to design a decentralized access protocol that minimizes the network AAoI, 
\begin{equation}
\Delta \triangleq \frac{1}{N}\sum_{n=1}^N {\Delta}_n.
\end{equation}

\subsection{Random Access Protocol}
\label{Sec: Decentralized Access Protocol}
We define the age gain of device $n$ at the beginning of slot $t$ as
\begin{align}
	\label{Eq: g}
g_{n,t} \triangleq h_{n,t} - w_{n,t},
\end{align}
which quantifies the reduction in instantaneous AoI upon a successful transmission of device $n$.
Based on the fact that $h_{n,t} \geq w_{n,t}$, it is clear that $g_{n,t} \geq 0$.

Following~\cite{Chen2022TIT}, we require each device $n$ with a non-empty buffer (i.e., $g_{n,t}\geq1$) to send its update according to the following T-AGDSA protocol, that is,
\begin{enumerate}
\item transmits at the beginning of slot $t$ with the probability $p_t\in(0,1]$ if $g_{n,t} \geq \Gamma_t$ where the threshold $\Gamma_t$ can be an arbitrary positive integer,
\item otherwise keeps silent at slot $t$.
\end{enumerate}
A device $n$ is said to be active in slot $t$ if $g_{n,t} \geq \Gamma_t$.

Then, we consider the following two versions of T-AGDSA with different settings of $\Gamma_t$ and $p_t$:
\begin{enumerate}
\item Basic T-AGDSA: for simple implementation~\cite{Yavascan2021JSAC}, the values of $\Gamma_t$ and $p_t$ are ﬁxed to $\Gamma^{\text{sta}}$ and $p^{\text{sta}}$, respectively, for all slots $t$.
\item Enhanced T-AGDSA: at the beginning of each slot $t$, each device determines the values of $\Gamma_t$ and $p_t$ based on the knowledge of age gains of all the devices.
Such knowledge comes from the globally available information, including the network parameters $N$, $\lambda$, $D$, previous channel status $c_{t-1}$, and previous access parameters $\Gamma_{t-1}$, $p_{t-1}$.
Note that, since the devices only use global information to compute $\Gamma_t$ and $p_t$, each device will obtain the same
values of them.
\end{enumerate}
It will be shown in Sections~\ref{sec:SAGRA}--\ref{Sec:Numerical Results} that basic T-AGDSA can be designed offline through theoretical modeling and is simpler to implement compared to enhanced T-AGDSA, but \textit{cannot} utilize the knowledge of age gains of all the devices to improve the AoI performance as done in enhanced T-AGDSA.

\begin{figure}[!htbp]
	\centering
	\includegraphics[width=3.3in]{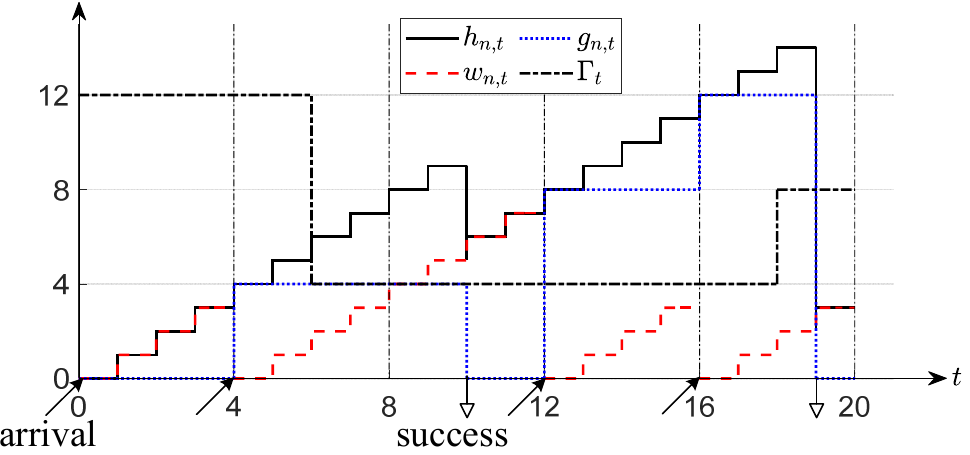}
	\caption{
 An example of $h_{n,t}$, $w_{n,t}$, $g_{n,t}$ and $\Gamma_t$ evolving over time under enhanced T-AGDSA when $D=4$. 
The black solid arrowhead indicates the arrival of an update,
whereas the black hollow arrowhead indicates the successful transmission of the most recently generated update of device $n$ in the corresponding slot.
}
	\label{fig:aoi}
\end{figure}

An example of $h_{n,t}$, $w_{n,t}$, $g_{n,t}$ and $\Gamma_t$ evolving over time under enhanced T-AGDSA when $D=4$ is shown in Fig.~\ref{fig:aoi}.

\subsection{Lower bound}
When $D=1$,~\cite{Chen2022TIT} derived a lower bound on the achievable network AAoI by assuming that all updates can be delivered instantaneously upon their arrival, without experiencing collisions.
We extend this bound to the case $D \geq 1$.
This bound is tighter when $N\lambda/D$ is smaller, which will be verified in Section~\ref{Sec:Numerical Results}.
The proof is given in Appendix.

\emph{Proposition 1}: For any transmission scheme under the system model specified in Section~\ref{sec: network model},
\begin{equation}\label{Eq: lowernound}
    \Delta \geq D/\lambda + (1-D)/2.
\end{equation}

\section{Modeling and Design of Basic T-AGDSA}  \label{sec:SAGRA}
In this section, we provide an analytical modeling approach to evaluate the network AAoI of basic T-AGDSA, and use this modeling to obtain optimal values of fixed threshold $\Gamma^{\text{sta}}$ and fixed transmission probability $p^{\text{sta}}$.

The symmetric scenario described in Section~\ref{sec:SystemModel} allows us to analyze the AAoI of an arbitrarily tagged device to represent the network AAoI.  
So, we omit the device index for analysis simplicity. 
To reflect the impact of the frame length $D$ better, we identify a slot $t$ by the tuple $(m, \nu)$, where $m=\lfloor t/D \rfloor$ and $\nu=t-mD$, for arbitrary $t\geq 0$. 
The main notations used in our analysis are listed in Table~\ref{table:notations}.

\begin{figure*}[!htbp]
	\centering
	\includegraphics[width=6.5in]{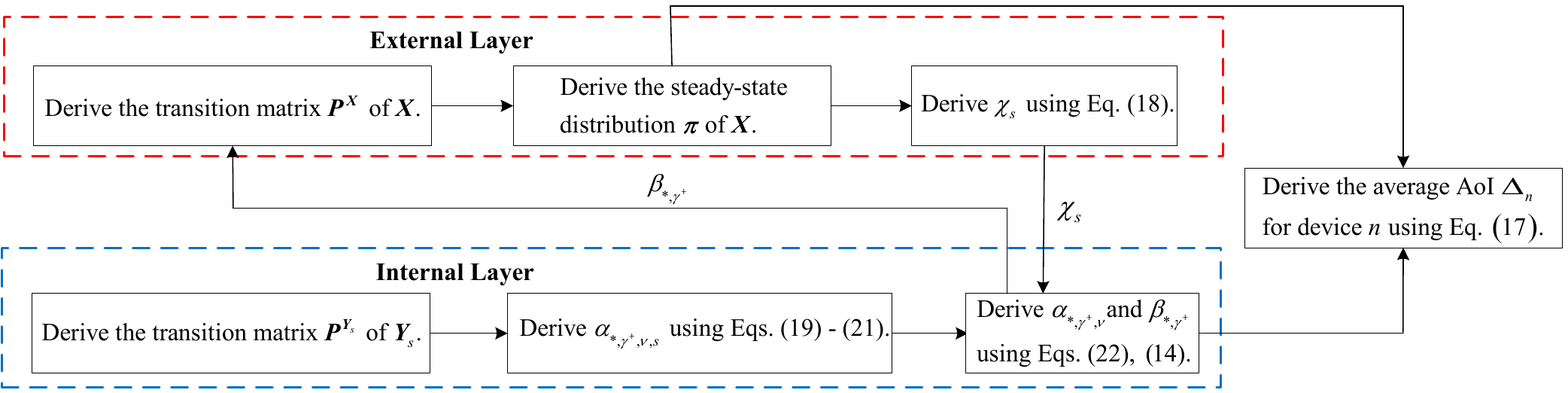}
	\caption{
 Flowchart illustrating the basic idea of analyzing basic T-AGDSA.
 }
	\label{fig: basic idea}
\end{figure*}

To calculate the AAoI of the tagged device, as shown in Fig.~\ref{fig: basic idea}, we adopt a multi-layer Markov model where the external layer manages the jumps between the beginnings of frames, while the internal layer manages the evolution during an arbitrary frame.
In the rest of this section, we explore how to establish these two coupled layers.

\begin{table}[!htbp]
\centering
\caption{Main notations used in Section III.}
\begin{tabular}{|c|c|} \hline
\textbf{Notation} & \textbf{Description} \\ \hline
$H_m, W_m, G_m$ & \makecell{The instantaneous AoI, local age, and age gain \\
                            of the tagged device at the beginning of frame $m$.}  \\ \hline
$\bm{X}$        & \makecell{An external DTMC with the infinite state space \\
                            $\mathcal{X}\triangleq\{(lD,kD)|l,k\in \mathbb{N}\}$.}  \\ \hline
$\bm{P}^{\bm{X}}$ & \makecell{The transition matrix of $\bm{X}$.} \\ \hline
$\alpha_{l,k,\nu}$ & \makecell{The probability that the tagged device \\transmits its $m$-th update
                            successfully \\in slot $(m,\nu)$ given $X_m = (lD,kD)$}.             \\ \hline
$\beta_{l,k}$   & \makecell{The probability that the tagged device \\transmits its $m$-th update
                            successfully \\in frame $m$ given $X_m = (lD,kD)$. }                 \\ \hline
$\alpha_{*,\gamma^+,\nu}$  & \makecell{$\alpha_{l,k,\nu}$ for $l \geq 0$ and $k \geq \gamma$.}    \\ \hline
$\beta_{*,\gamma^+}$ & \makecell{$\beta_{l,k}$ for $l \geq 0$ and $k \geq \gamma$.}    \\ \hline
$\pi_{lD,kD}$   & \makecell{The steady-state probability of $\bm{X}$ \\
                            staying at state $(lD,kD)$ for $l,k\in\mathbb{N}$.}\\ \hline
$S_m$  & \makecell{The number of active devices \\not including the tagged device
                   \\at the beginning of an arbitrary frame $m$. }               \\ \hline
$\chi_{s}$ & \makecell{The probability mass function of $S_m = s$.}         \\ \hline
$\bm{Y}_s$ & \makecell{An absorbing DTMC with the finite state space \\
                       $\mathcal{Y}_s\triangleq\{0,1,\ldots,s,suc\}$.} \\ \hline
$Y_{s,\nu}$ & \makecell{State indicating the transmission results of \\all devices before the beginning of slot $(m,\nu)$.} \\ \hline
$\bm{P}^{\bm{Y_s}}$ & \makecell{The transition matrix of $\bm{Y}_s$.} \\ \hline
$\alpha_{*,\gamma^+,\nu,s}$ & \makecell{The probability that the tagged device \\transmits its $m$-th update 
                                       successfully \\at slot $(m,\nu)$ when $G_m \geq \gamma D$, $S_m = s$.}\\ \hline
\end{tabular}
\label{table:notations}
\end{table}

\subsection{External Layer} \label{sec:External Layer}
Let $W_m$ and $H_m$ denote the local age and instantaneous AoI of the tagged device at the beginning of frame $m$, respectively.
By Eq.~\eqref{eq:Evolution_w} and the traffic pattern described in Section~\ref{sec:SystemModel}, the evolution of $W_m$ with $W_0 = 0$ can be expressed as
\begin{equation}
    \label{eq:deqn_W}
	W_{m+1}\! =\!
	\begin{cases}
		0,    & \text{if an update arrives at the} \\
              & \text{beginning of frame $m+1$,}\\
		W_m + D,  & \text{otherwise.}
	\end{cases}
\end{equation}
By Eq.~\eqref{eq:Evolution_h}, the evolution of $H_m$ with $H_0 = 0$ can be expressed as
\begin{equation}
    \label{eq:deqn_H}
	H_{m+1}\! =\!
	\begin{cases}
		W_m + D,    & \text{if an update is successfully} \\
                    & \text{transmitted during frame $m$,} \\
		H_m + D,  & \text{otherwise.}
	\end{cases}
\end{equation}
Denote the age gain of the tagged device at the beginning of frame $m$ by $G_m$. 
By Eq. \eqref{Eq: g}, we have
\begin{align}\label{eq: deqn_G}
	G_m = H_m-W_m.
\end{align}

Consider a state process $\bm{X} \triangleq \{X_m, m \in \mathbb{N}\}$ where $X_m \triangleq (W_m, G_m)$.
By Eqs.~\eqref{eq:deqn_W}--\eqref{eq: deqn_G}, we observe that the transition to the next state in $\bm{X}$ depends only on the present state and not on the previous states.
Hence, $\bm{X}$ can be viewed as a DTMC with the infinite state space $\mathcal{X}\triangleq\{(lD,kD)|l,k\in \mathbb{N}\}$.

For an arbitrary frame $m$ with $X_m = (lD,kD)$, let $\alpha_{l,k,\nu}$ and $\beta_{l,k}$ denote the probabilities that the tagged device transmits its update successfully at slot $(m,\nu)$ and in frame $m$, respectively.
Obviously, $\beta_{l,k} = \sum_{\nu = 0}^{D-1} \alpha_{l,k,\nu}$.
According to the evolution of $W_m$ and $G_m$ given in Eqs.~\eqref{eq:deqn_W}--\eqref{eq: deqn_G}, the state transition probabilities of $\bm{X}$ can be obtained as
\begin{align}
&P^{\bm{X}}_{(lD,kD),(l'D,k'D)} \notag \\
&\triangleq 
 \text{Pr}\big(X_{m+1} = (l'D,k'D) \mid X_{m} = (lD,kD)\big) \notag \\
        &=
	\begin{cases} 
		\lambda\beta_{l,k},      & \text{if } l'=0,k'=l+1, \\
		\lambda(1 - \beta_{l,k}),  & \text{if } l'=0,k'=l+k+1, \\
        (1-\lambda)\beta_{l,k},  & 
        \begin{aligned}
         \text{if }  l'& = l+1, k' = 0, 
        \end{aligned}\\
        (1-\lambda)(1-\beta_{l,k}),& \begin{aligned}
        \text{if } l'& = l+1, k' = k, 
        \end{aligned}\\
            0,              & \text{otherwise.}
	\end{cases}
 \label{eq: Xtransition}
\end{align}

\begin{figure}[!htbp]
  \centering
    \subfloat[$l \geq 0, 0 \leq k \leq \gamma-1.$]
      {     
      \label{fig:subfig5}\includegraphics[width=2.5in]{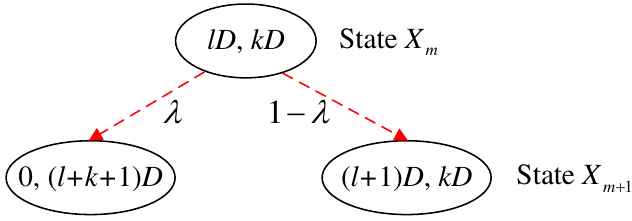}
      }
      \\
     \subfloat[$l \geq 0, k \geq \gamma.$]
      {
      \label{fig:subfig6}\includegraphics[width=3.3in]{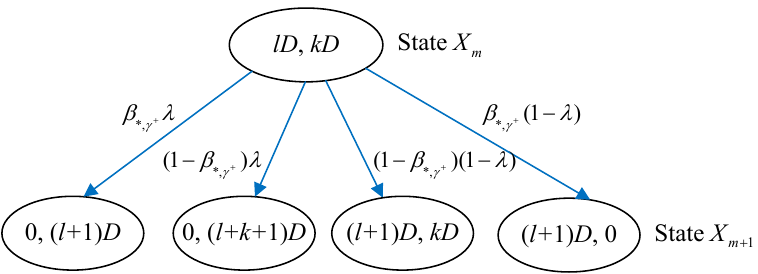}
      }
 \caption{The external DTMC $\bm{X}$.}
  \label{fig:DTMC_case}
\end{figure}

Consider that under the traffic pattern described in Section~\ref{sec:SystemModel}, 
the age gain of the tagged device during frame $m$ remains unchanged if the tagged device fails to transmit its update successfully in frame $m$; 
otherwise, the age gain will reduce to zero and remain zero in the subsequent slots during frame $m$ after a successful transmission. 
Thus, the age gain of the tagged device in each slot takes value from $\{0, D, 2D, \ldots\}$. 
This observation allows us to set $\gamma \triangleq \lceil \Gamma^{\text{sta}}/D \rceil$, 
and discuss possible values of $\alpha_{l,k,\nu}$ and $\beta_{l,k}$ in Eq.~\eqref{eq: Xtransition} based on different values of $l$, $k$, $\nu$ and $\gamma$.

\underline{\emph{Case 1}:} When $l \geq 0, 0 \leq k \leq \gamma-1$, we have $G_m \leq (\gamma-1)D < \Gamma^{\text{sta}}$. 
Consider that the tagged device keeps its age gain unchanged during frame $m$ if it does not make a successful transmission during frame $m$. 
So, the tagged device always keeps silent in frame $m$ as its age gain is always smaller than $\Gamma^{\text{sta}}$. 
Then we obtain 
\begin{align}
\alpha_{l,k,\nu} &= 0, \label{eq: case1}\\
\beta_{l,k} &= \sum_{\nu = 0}^{D-1} \alpha_{l,k,\nu} = 0,\label{eq: case11}
\end{align}
if $l \geq 0, 0\leq k \leq \gamma-1$ and $0\leq \nu \leq D-1$.
With Eqs.~\eqref{eq: Xtransition}--\eqref{eq: case11}, the state transitions for this case are illustrated in Fig.~\ref{fig:DTMC_case}(a).

\underline{\emph{Case 2}:} When $l \geq 0, k \geq \gamma$, we have $G_m \geq \gamma D \geq \Gamma^{\text{sta}}$, the tagged device transmits its update with a fixed probability $p^{\text{sta}}$ at the beginning of slot $(m,\nu)$ when $0\leq \nu \leq D-1$ until a successful transmission. 
Note that the tagged device behaves the same during frame $m$ regardless of the values of $W_m$, $G_m$ when $G_m \geq \gamma D$.
So, we rewrite $\alpha_{l,k,\nu}$ and $\beta_{l,k}$ simply as
\begin{align}
\alpha_{l,k,\nu} &= \alpha_{*,\gamma^+,\nu}, \label{eq: case2}\\
\beta_{l,k} &= \sum_{\nu = 0}^{D-1} \alpha_{*,\gamma^+,\nu} = \beta_{*,\gamma^+},
\label{eq: case22}
\end{align}
if $l \geq 0, k \geq \gamma$ and $0\leq \nu \leq D-1$.
Note that $\alpha_{*,\gamma^+,\nu}$ and $\beta_{*,\gamma^+}$ are independent of values of $l$ and $k$.
With Eqs.~\eqref{eq: Xtransition},~\eqref{eq: case2}, and~\eqref{eq: case22}, the state transitions for this case is illustrated in Fig.~\ref{fig:DTMC_case}(b).

Note that the state $(0,0)$ in $\bm{X}$ is an ephemeral state only occurring when $m = 0$ (i.e., $t = 0$), while the remaining states are all in the same recurrent class and occur when $m\geq 1$ (i.e., $t \geq D$).
As $m$ increases, $\bm{X}$ will get absorbed in the recurrent class, and it will stay there forever. 
Denote by $\bm{\pi} \triangleq (\pi_{lD,kD})_{l,k \in \mathbb{N}}$ the steady-state distribution of $\bm{X}$.
Each element $\pi_{lD,kD}$ denotes the steady-state probability of $\bm{X}$ staying at state $(lD,kD)$.
We assume that $\pi_{0,0}=0$.

Then, for different states in $\bm{X}$, we consider the following two cases for evaluating the AAoI of the tagged device during an arbitrary frame $m$ with $X_m=(lD,kD)$.

\underline{\emph{Case 1}:} The tagged device transmits its update successfully at slot $(m,\nu)$ given $X_m=(lD,kD)$.
Let ${\Delta_{l,k,\nu}}$ denote the AAoI of the tagged device during frame $m$ when this event occurs. 
We have
\begin{align}
	{\Delta_{l,k,\nu}}&=\frac{1}{D}\Big(\sum_{\nu'=0}^{\nu}\big((l+k)D+\nu'\big)+\sum_{\nu'=\nu+1}^{D-1}(lD+\nu')\Big)  \notag \\
         &=lD + k(\nu+1) + \frac{D-1}{2}.  \label{eq:Delta1h}
\end{align}

\underline{\emph{Case 2}:} The tagged device fails to transmit its update successfully during frame $m$ given $X_m=(lD,kD)$. 
Let ${\Delta_{l,k,*}}$ denote the AAoI of the tagged device during frame $m$ when this event occurs. 
We have
\begin{align}
	{\Delta_{l,k,*}}=\frac{\sum_{\nu=0}^{D-1}\big((l+k)D+\nu\big)}{D}=(l+k)D+\frac{D-1}{2}.  \label{eq:Delta1*}
\end{align}

Based on the DTMC $\bm{X}$ and Eqs.~\eqref{eq:Delta1h},~\eqref{eq:Delta1*}, the AAoI of an arbitrary device $n$ can be derived as
\begin{align}
  {\Delta_n}
  &=\sum_{l=0}^{\infty}\sum_{k=0}^{\infty}\pi_{lD,kD}\big(\sum_{\nu=0}^{D-1}\alpha_{l,k,\nu}{\Delta_{l,k,\nu}}\!+\!(1-\beta_{l,k}) \Delta_{l,k,*} \big) \notag \\
   &= \!\sum_{l=0}^{\infty}\sum_{k=\gamma}^{\infty}\pi_{lD,kD}\big(\sum_{\nu=0}^{D-1}\alpha_{*,\gamma^+,\nu}{\Delta_{l,k,\nu}}
	          \!+\! (1\!-\!\beta_{*,\gamma^+})\Delta_{l,k,*}\big) \notag\\
    &\quad + \sum_{l=0}^{\infty}\sum_{k=0}^{\gamma-1}\pi_{lD,kD}\Delta_{l,k,*}.     \label{eq:Delta derive}
\end{align}

In the following, we will derive $\alpha_{*,\gamma^+,\nu}$, which is necessary to compute ${\Delta_n}$ based on Eqs.~\eqref{eq:Delta1h}--\eqref{eq:Delta derive}.

\subsection{Internal Layer to Evaluate $\alpha_{*,\gamma^+,\nu}$}

Note that whether the tagged device can transmit its update successfully depends on the behaviors of all active devices during frame $m$. 
Let a random variable $S_m$ denote the number of active devices not including the tagged device at the beginning of an arbitrary frame $m$, and let $\chi_s$ denote the probability mass function of $S_m = s$.
Following~\cite {chen2020age}, we make a simplifying decoupling assumption that the states of all the devices are independent of each other.
Then, based on the binomial distribution, we have
\begin{align}
\chi_s=\binom{N-1}{s}\Big(\sum_{l=0}^{\infty}\sum_{k=\gamma}^{\infty}\pi_{lD,kD}\Big)^{s}\Big(\sum_{l=0}^{\infty}\sum_{k=0}^{\gamma-1}\pi_{lD,kD}\Big)^{N-1-s},
  \label{eq:Pr_s1s2}
\end{align}
for each $0\leq s \leq N-1$.

\begin{figure}[!htbp]
	\centering
	\includegraphics[width=3in]{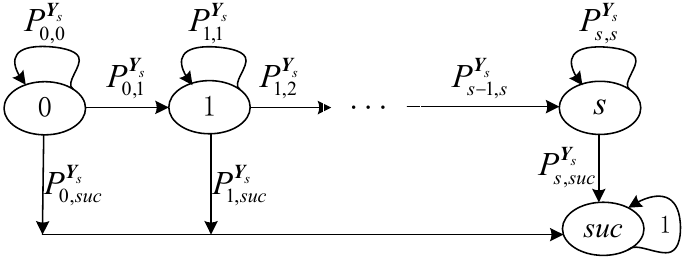}
	\caption{The internal absorbing DTMC $\bm{Y}$.}
	\label{fig:DTMC Y}
\end{figure}

Consider an arbitrary frame $m$ with $G_m \geq \gamma D$ and $S_m = s$.
Define $\bm{Y}_s \triangleq \{ Y_{s,\nu}, \nu=0,1,\ldots,D \}$ as an absorbing DTMC with the finite state space $\mathcal{Y}_s\triangleq\{0,1,\ldots,s,suc\}$, as shown in Fig.~\ref{fig:DTMC Y}.
The states $Y_{s,\nu}=y$ with $0\leq \nu \leq D, 0\leq y \leq s$ are transient states indicating that, during frame $m$, the tagged device has not transmitted successfully before the beginning of slot $(m,\nu)$ while other $y$ devices have transmitted successfully before the beginning of slot $(m,\nu)$.
The state $Y_{s,\nu}=suc$ is an absorbing state indicating that, during frame $m$, the tagged device has transmitted successfully before the beginning of slot $(m,\nu)$.
For convenience, the slot index $(m, D)$ is used to denote the slot index $(m+1, 0)$ here. 
As shown in Fig.~\ref{fig:DTMC Y}, the state transition probabilities of $\bm{Y}_s$ can be obtained as
\begin{align} 
\label{Y_transition}
   &P^{\bm{Y}_s}_{y,y'} \triangleq \mathrm{Pr} (Y_{s,\nu+1}=y' \mid Y_{s,\nu}=y) \notag \\ 
    =& 
    \begin{cases}
        1 \!-\! (s\!-\!y\!+\!1)p^{\text{sta}}(1\!-\!p^{\text{sta}})^{s-y},&\text{if } 0\le y\le s, y' = y, \\
        (s-y)p^{\text{sta}}(1-p^{\text{sta}})^{s-y},  &\text{if } 0\le y\le s - 1,\\ 
                                                       &\quad y' = y+1, \\
        p^{\text{sta}}(1-p^{\text{sta}})^{s-y}, &\text{if } 0\le y\le s, y' = suc, \\
        1, &\text{if } y = y' = suc, \\
        0, &\text{otherwise}.
    \end{cases}
\end{align}
The first case in Eq.~\eqref{Y_transition} corresponds to that no device makes a successful transmission in slot $(m, \nu)$ when $0\le y\le s$.
The second case corresponds to that one of the other $(s-y)$ devices makes a successful transmission in slot $(m, \nu)$ when $0\le y\le s - 1$.
The third case corresponds to that the tagged device makes a successful transmission in slot $(m, \nu)$ when $0\le y\le s$.
The fourth case corresponds to that the tagged device has made a successful transmission before the beginning of slot $(m, \nu)$.

Let $\bm{\varphi}_{s,\nu}$ denote the state vector of $Y_{s,\nu}$, where the $i$-th element corresponds to the state $i-1$ for $1\leq i \leq s+1$ and the last element corresponds to the state $suc$.
Then, given the priori state vector $\bm{\varphi}_{s,0} \triangleq [1,0,0,\ldots,0]$ and the transition matrix $\bm{P}^{\bm{Y}_s}$ based on Eq.~\eqref{Y_transition}, by applying a simple power method, we have
\begin{align}
		\bm{\varphi}_{s,\nu}&=\bm{\varphi}_{s,0} {(\bm{P}^{\bm{Y}_s})}^{\nu}, 
  \label{eq: varphi}
\end{align}
for each $0\leq \nu \leq D$ and $0\leq s \leq N-1$.

Let $\alpha_{*,\gamma^+,\nu,s}$ denote the probability that the tagged device transmits successfully in slot $(m,\nu)$ of an arbitrary frame $m$ with $G_m \geq \gamma D$, $S_m = s$. 
Then, we have
\begin{align}
\alpha_{*,\gamma^+,\nu,s}&=\bm{\varphi}_{s,\nu+1}(s+2)-\bm{\varphi}_{s,\nu}(s+2),\label{eq:alpha_s}
\end{align}
for each $0\leq \nu \leq D-1$ and $0\leq s \leq N-1$.
By Eqs.~\eqref{eq:Pr_s1s2} and \eqref{eq:alpha_s}, we have
\begin{align}
	\alpha_{*,\gamma^+,\nu}
			& =\sum_{s=0}^{N-1}\chi_s\alpha_{*,\gamma^+,\nu,s},
   \label{eq:alpha_kh}
\end{align}
for each $0\leq \nu \leq D-1$.

\subsection{Evaluation of $\Delta_n$}
Now we are ready to use the following three steps to compute $\Delta_n$ by connecting the external and internal layers proposed in previous subsections.

\underline{\emph{Step 1}:}
Based on the transition probabilities in Eq.~\eqref{eq: Xtransition}, the steady-state distribution $\bm{\pi}$ can be obtained by solving a set of linear equations
\begin{align}
\label{pi}
    \bm{\pi}\bm{P}^{\bm{X}} = \bm{\pi},
\end{align}
and the normalizing condition
\begin{align}
\label{normalizing}
\sum_{l=0}^{\infty}\sum_{k=0}^{\infty}\pi_{lD,kD}=1.
\end{align}
Since $\beta_{*,\gamma^+}$ is involved as the only unknown parameter in the transition matrix $\bm{P}^{\bm{X}}$, each $\pi_{lD,kD}$ can be expressed as a function of $\beta_{*,\gamma^+}$ using a mathematical induction method based on Eqs.~\eqref{eq: Xtransition},~\eqref{pi},~\eqref{normalizing}.
Meanwhile, $\alpha_{*,\gamma^+,\nu}$ in Eq.~\eqref{eq: case22} can be expressed as a function of $\pi_{lD,kD}, l,k\in\mathbb{N}$ based on Eqs.~\eqref{eq:Pr_s1s2}--\eqref{eq:alpha_kh}.
Hence, we can obtain the value of $\beta_{*,\gamma^+}$ by solving Eq.~\eqref{eq: case22} using numerical methods like the fixed-point iteration method. 

\underline{\emph{Step 2}:}
With the value of $\beta_{*,\gamma^+}$, we can obtain the values of $\pi_{lD,kD}, l,k\in\mathbb{N}$ by Eqs.~\eqref{eq: Xtransition},~\eqref{pi} and~\eqref{normalizing}, and then obtain the values of $\alpha_{*,\gamma^+,\nu}, 0\leq \nu \leq D-1$ by Eqs.~\eqref{eq:Pr_s1s2}--\eqref{eq:alpha_kh}.

\underline{\emph{Step 3}:}
With the values of $\pi_{lD,kD}, l,k\in\mathbb{N}$, $\alpha_{*,\gamma^+,\nu}, 0\leq \nu \leq D-1$ and $\beta_{*,\gamma^+}$, we can obtain the AAoI for an arbitrary device $n$ by Eq.~\eqref{eq:Delta derive}.

\vspace{2pt}
\noindent \emph{Remark 1:} 
When $\Gamma^{\text{sta}} = \lambda = 1$, we note that we can drop the subscripts $l,k$ of $\beta_{l,k}$ and have $\pi_{0,kD}=\beta(1-\beta)^{k-1}$ for $k \geq 1$ since $\beta_{l,k}$ is independent of $l$ and $k$.
So, our modeling approach is reduced to that in~\cite{bae2022age}. 

\vspace{2pt}
\noindent \emph{Remark 2:} 
When $D=\lambda=1$ (i.e., the GAW traffic), we have $W_m = 0$, $H_m = G_m$ for each $m\in \mathbb{N}$, and
\begin{align}
	\beta_{*,\gamma} = \beta_{*,\gamma^+} = \alpha_{*,\gamma,0} 
        &= \alpha_{*,\gamma^+,0} \notag \\
        &=\sum_{s=0}^{N-1}\chi_s p^{\text{sta}} (1-p^{\text{sta}})^s\notag\\
	&=p^{\text{sta}}\Big(1-p^{\text{sta}}\sum_{k=\gamma}^{\infty}\pi_{0,k}\Big)^{N-1}. 
\end{align}
So, our modeling approach is reduced to that in~\cite{chen2020age}. 

\vspace{2pt}
\noindent \emph{Remark 3:}
In general, there may exist multiple solutions of $\beta_{*,\gamma^+}$ in step 1 since Eq.~\eqref{eq:Pr_s1s2} assumes that the states
of all the devices are independent of each other, which requires simulations to help us identify the correct solution.   
When $D=\lambda=1$,~\cite{Yavascan2021JSAC} presented a precise steady-state analysis without this ideal assumption.
A key idea therein is to utilize an inherent feature for $D=\lambda=1$, that is, different inactive devices have different age gains.   
However, it is inapplicable for $D>1$ or $\lambda<1$ because there may exist multiple inactive devices with the same age gain.

\subsection{Seeking Optimal $\Gamma^{\text{sta}}$ and $p^{\text{sta}}$ }\label{sec: seeking optimal value}
In order to seek optimal values of $\Gamma^{\text{sta}}$ and $p^{\text{sta}}$, we can view $\Delta_n$ given in Eq.~\eqref{eq:Delta derive} as a function
of both $\Gamma^{\text{sta}}$ and $p^{\text{sta}}$, denoted by $\Delta_n(\Gamma^{\text{sta}}, p^{\text{sta}})$.
However, it is difficult to obtain the gradient of $\Delta_n(\Gamma^{\text{sta}}, p^{\text{sta}})$ due to the lack of an explicit expression.
So, well-known two-dimensional gradient-free search methods, such as the Hooke-Jeeves method, and Rosenbrock method, can be applied.
Moreover, since the age gain of each device is always an integral multiple of $D$ as described in Section~\ref{sec:External Layer}, we can consider only $\Gamma^{\text{sta}} = kD, k = 1,2,\ldots$ to reduce the search space.

\section{Design of Enhanced T-AGDSA}  
\label{sec: A practical Scheme}

In this section, we propose an enhanced T-AGDSA scheme that allows each device to adjust the threshold $\Gamma_t$ and the transmission probability $p_t$ for maximizing the estimated network EAR per slot, based on the knowledge of age gains of all the devices.
After introducing the basic idea of our design, we will present a comprehensive explanation of the three key steps as shown in Fig.~\ref{fig: basic idea2}.

\subsection{Basic Idea}
Define the AoI reduction of device $n$ in slot $t$ as
\begin{align}\label{eq: rnt}
    r_{n,t} \triangleq h_{n,t} - h_{n,t+1}.
\end{align}
We can use Eqs.~\eqref{eq:Evolution_h},~\eqref{Eq: g} and~\eqref{eq: rnt} to compute $r_{n,t}$ as follows.
\begin{align}\label{eq: rnt_s}
    r_{n,t} = 
    \begin{cases}
        (w_{n,t} \!+\! g_{n,t}) \!-\! (w_{n,t} \!+\! 1) \\
        = g_{n,t} - 1, &\text{if device $n$ successfully}\\&\text{transmits in slot $t$,}\\
        h_{n,t} - (h_{n,t} + 1)
            = -1, &\text{otherwise.}\\
    \end{cases}
\end{align}
Let $\theta_{n,t}$ denote the success probability when device $n$ transmits in slot $t$.
From Eq.~\eqref{eq: rnt_s}, we can obtain the EAR of device $n$ in slot $t$ as
\begin{align}\label{eq: Rnt}
    R_{n,t} \triangleq \mathbb{E}(r_{n,t}) 
        & = 
        \big((g_{n,t} - 1) p_t\theta_{n,t} - (1- p_t\theta_{n,t})\big)I_{g_{n,t}\geq\Gamma_t} \notag\\
        &\quad + (-1)I_{0\leq g_{n,t}<\Gamma_t} \notag \\
            &= \big(g_{n,t} p_t\theta_{n,t}-1\big)I_{g_{n,t}\geq\Gamma_t} - I_{0\leq g_{n,t}<\Gamma_t}  \notag \\
        &= -1 + g_{n,t} p_t\theta_{n,t}I_{g_{n,t}\geq\Gamma_t},\notag \\
    &= -1 + g_{n,t}p_t(1-p_t)^{  u_{n,t}}I_{g_{n,t}\geq\Gamma_t}, 
\end{align}
where $I_{\psi}$ is the indicator function of event $\psi$ and $u_{n,t} \triangleq \sum_{n'=1,n'\neq n}^N I_{g_{n',t}\geq\Gamma_t}$ denotes the number of active devices not including device $n$ in slot $t$.
Then, the network EAR in slot $t$ can be obtained as
\begin{align}\label{eq: rt}
    R_t \triangleq \frac{\sum_{n=1}^{N}R_{n,t}}{N}.
\end{align}

This section will investigate how each device uses globally available information to choose $\Gamma_t$ and $p_t$ for maximizing $R_t$.
Note that, although it may be possible to achieve better results by using information local to the devices, or using a strategy to maximize the long-run network EAR that is not ``one-step look-ahead'', we do not pursue these possibilities here.

From Eqs.~\eqref{eq: Rnt} and \eqref{eq: rt}, it is easy to see that the knowledge of the age gains $g_{n,t}$, $n\in \mathcal{N}$ is essential for maximizing $R_t$.
However, in practice, it is impossible for each device to obtain precise values of the age gains of other devices. 
So, we require each device to keep a posteriori joint probability distribution of local age and age gain of an arbitrarily tagged device at the beginning of slot $t$, given all of the globally available information.
We denote such a distribution by $(\hat{f}_{t,w,g})_{w,g\in\mathbb{N}}$,
where $\hat{f}_{t,w,g} \triangleq \text{Pr}(w_{n,t} = w, g_{n,t} = g), w,g\in\mathbb{N}$.
Based on $(\hat{f}_{t,w,g})_{w,g\in\mathbb{N}}$, a brief introduction of the three key steps of the proposed enhanced T-AGDSA as shown in Fig.~\ref{fig: basic idea2} is given below:

\underline{\emph{Step 1}:}
At the beginning of slot $t$, each device uses $(\hat{f}_{t,w,g})_{w,g\in\mathbb{N}}$ to obtain an estimate of network EAR, denoted by $\hat{R}_t$, and then chooses $\Gamma_t$ and $p_t$ that maximize $\hat{R}_t$.

\underline{\emph{Step 2}:}
At the end of slot $t$, each device uses the observed channel status $c_t$ to update 
$(\hat{f}_{t,w,g})_{w,g\in\mathbb{N}}$ in a Bayesian manner.
We denote the resulting distribution in this step by $(\hat{f}_{t^+,w,g})_{w,g\in\mathbb{N}}$, where $t^+$ represents the end of slot $t$.

\underline{\emph{Step 3}:}
At the beginning of slot $t+1$, if $t = mD-1, m \in \mathbb{Z}^+$, each device obtains $(\hat{f}_{t+1,w,g})_{w,g\in\mathbb{N}}$ using $(\hat{f}_{t^+,w,g})_{w,g\in\mathbb{N}}$ and the update generation probability $\lambda$,
otherwise, each device obtains $(\hat{f}_{t+1,w,g})_{w,g\in\mathbb{N}} = (\hat{f}_{t^+,w,g})_{w,g\in\mathbb{N}}$.

\begin{figure*}[!htbp]
	\centering
	\includegraphics[width=6in]{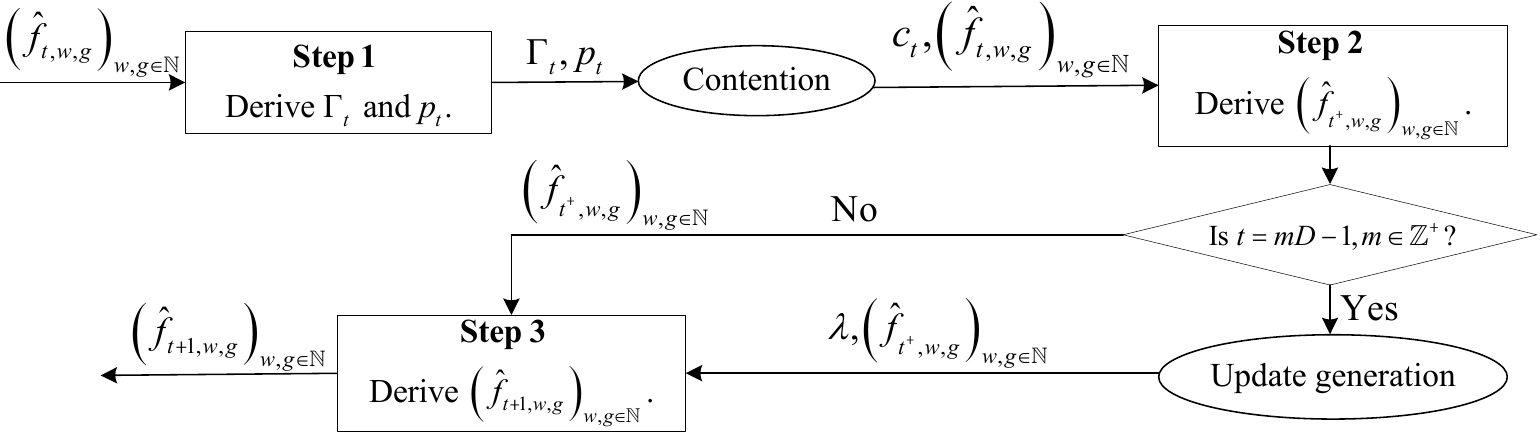}
	\caption{Flowchart illustrating the basic idea of the proposed enhanced T-AGDSA.}
	\label{fig: basic idea2}
\end{figure*}

\vspace{2pt}
\noindent \emph{Remark 4:}
In AAT~\cite{Chen2022TIT}, $\Gamma_t$ is chosen so that the effective sum arrival rate approaches $1/e$ as close as possible and $p_t$ is then chosen for maximizing the instantaneous network throughput.
However, such a setting may yield unsatisfactory network EAR, since it allows the devices with low age gains to compete for the transmission opportunity as soon as the effective sum arrival rate does not exceed $1/e$.  
In other words, in a slot, higher network throughput {\em{cannot}} be certainly converted to larger network EAR.

\vspace{2pt}
\noindent \emph{Remark 5:}
In practical T-DFSA~\cite{Moradian2024TCOM}, the maximum value of the thresholds that make the estimated expected number of active devices not smaller than a certain number (searched by simulations) is chosen.
However, such a setting may yield unsatisfactory network EAR, since its design objective may be far from maximizing the network EAR, especially when the probability distribution of the estimated number of active devices is divergent.

\vspace{2pt}
\noindent \emph{Remark 6:}
Both AAT~\cite{Chen2022TIT} and practical T-DFSA~\cite{Moradian2024TCOM} ideally assume that the age gain and local age of a device are independent of each other, thus only consider the distribution of age gain.
However, there is a strong dependency between them as described in Eq.~\eqref{Eq: g}.
This is indeed why we consider $(\hat{f}_{t,w,g})_{w,g\in\mathbb{N}}$.

\vspace{2pt}
\noindent \emph{Remark 7:}
The AAT~\cite{Chen2022TIT} utilizes only the collision feedback to update its estimate, while our enhanced T-AGDSA scheme utilizes the ternary feedback (idle, success, collision).
Nevertheless, our scheme requires no additional overhead owing to the ACK mechanism.

\subsection{Choosing $\Gamma_t$ and $p_t$ based on $(\hat{f}_{t,w,g})_{w,g\in\mathbb{N}}$}

In the following, we present how to estimate the network EAR in slot $t$ based on $(\hat{f}_{t,w,g})_{w,g\in\mathbb{N}}$ by assuming that the states of all the devices are independent of each other.

Let $\hat{u}_{t}$ denote the estimated number of active devices not including an arbitrary device in slot $t$.
Based on the binomial distribution, we have
\begin{align}\label{eq: xi1}
\xi_{t,u} \triangleq \text{Pr}(\hat{u}_{t} = u) = \binom{N-1}{u}\rho_t^{u}(1-\rho_t)^{N-1-u},
\end{align}
for each $0\leq u \leq N-1$, where
\begin{align}
    \rho_t = \sum_{w=0}^{\infty}\sum_{g=\Gamma_t}^{\infty}\hat{f}_{t,w,g},
\end{align}
denotes the probability of an arbitrary device being active in slot $t$.

From Eq.~\eqref{eq: xi1}, we obtain the estimated success probability of an arbitrary transmission in slot $t$ as follows.
\begin{align}\label{eq: theta}
    \hat{\theta}_t = 
    \sum_{u=0}^{N-1}\xi_{t,u}(1-p_t)^{u}.
\end{align}
From Eqs.~\eqref{eq: rnt_s} and~\eqref{eq: theta}, we can obtain the following estimate of network EAR in slot $t$.
\begin{align}
    \hat{R}_{t} 
    &=  \sum_{w=0}^{\infty}\sum_{g=0}^{\infty}
    \hat{f}_{t,w,g}\big(-1 + g p_t\hat{\theta}_t I_{g\geq\Gamma_t}\big)\notag \\
    &= -1 + \Big(\sum_{w=0}^{\infty}\sum_{g=\Gamma_t}^{\infty}
    \hat{f}_{t,w,g}g\Big)p_t\hat{\theta}_t \notag \\
    &= -1 + \Big(\sum_{w=0}^{\infty}\sum_{g=\Gamma_t}^{\infty}
    \hat{f}_{t,w,g}g\Big)p_t\sum_{u=0}^{N-1}\xi_{t,u}(1-p_t)^{u}.\label{eq: eRt}
\end{align}

We can view $\hat{R}_t$ as a function of $\Gamma_t$ and $p_t$, denoted by $\hat{R}_t(\Gamma_t,p_t)$.
We see from Eq.~\eqref{eq: eRt} that, for each given $\Gamma_t\geq 1$, the maximization of $\hat{R}_t(\Gamma_t,p_t)$ is equivalent to the maximization of $p_t\hat{\theta}_t$.
So we can obtain the value $p_t^o$ that maximizes $p_t\hat{\theta}_t$ by differentiating and root finding since $p_t\hat{\theta}_t$ is a polynomial of $p_t$.
However, in practice, such computation would probably be excessive.
Following~\cite{rivest1987,Thilina2022}, we can approximate $p_t^o$ as follows.
\begin{align}\label{eq: pto}
    \hat{p_t^o} = \min \Big\{\frac{1} {\sum_{u'=1}^{N}\binom{N}{u'}\rho_t^{u'}(1-\rho_t)^{N-u'}u'},1\Big\}.
\end{align}
Since the function $\hat{R}_t(\Gamma_t,\hat{p_t^o})$ is unimodal in our implementation, we can obtain the value $\hat{\Gamma_t^o}$ that maximizes $\hat{R}_t(\Gamma_t,\hat{p_t^o})$ through an efﬁcient one-dimensional search method.
Moreover, due to the same argument in Section~\ref{sec: seeking optimal value}, we can consider only $\Gamma_t = kD, k = 1,2,\ldots$ to reduce the search space.

\subsection{Computing $(\hat{f}_{t^+,w,g})_{w,g\in\mathbb{N}}$ Using Channel Observations}
Let $w_{n,t^+}$ and $g_{n,t^+}$ denote the local age and age gain of an arbitrary device $n$ at the end of slot $t$, respectively. 
Given all globally available information at the end of slot $t$, each device is able to compute $\hat{f}_{t^+,w',g'}$ using Bayes' rule as follows.
\begin{align}
\label{eq: Bayesian update}
    &\hat{f}_{t^+,w',g'} \notag \\
    &\triangleq 
    \text{Pr}\big(w_{n,t^+} = w',g_{n,t^+}=g'|\Gamma_t,p_t,c_t=c,(\hat{f}_{t,w,g})_{w,g\in\mathbb{N}}\big) \notag \\
    & = \frac{1}{\rho}\sum_{w,g\in\mathbb{N}}\hat{f}_{t,w,g} \text{Pr} \Big(w_{n,t^+} = w',g_{n,t^+}=g',c_t=c|\Gamma_t, \notag \\
                                                  &\quad\quad\quad\quad\quad\quad\quad\quad\quad\quad\quad  p_t,w_{n,t} = w,g_{n,t}=g\Big),
\end{align}
for each $t,w',g'\in\mathbb{N}$,
where 
\begin{align}
    \rho 
    & = \sum_{w'',g''\in\mathbb{N}}\sum_{w,g\in\mathbb{N}}\hat{f}_{t,w,g}\text{Pr}\Big(w_{n,t^+} = w'',g_{n,t^+}=g'',\notag \\
    &\quad\quad\quad\quad\quad\quad c_t=c|\Gamma_t, p_t,w_{n,t} = w,g_{n,t}=g\Big),
\end{align}
and
    \begin{align}
    \label{eq: Bayesian update1}
        &\text{Pr}\big(w_{n,t^+} = w',g_{n,t^+} = g',c_t = c|\Gamma_t,p_t,w_{n,t} = w,g_{n,t} = g\big) \notag \\
        & = 
        \begin{cases}
            (1-p_t)\sum_{u=0}^{N-1}\xi_{t,u}(1-p_t)^{u},\\
            \quad\quad\quad\quad\quad \text{if }g\geq\Gamma_t,c = 0,w' = w+1,g' = g,\\
            \sum_{u=0}^{N-1}\xi_{t,u}(1-p_t)^{u},\\
            \quad\quad\quad\quad\quad\text{if }0\leq g <\Gamma_t,c = 0,w'\! = \!w+1,g' \!=\! g,\\
            (1-p_t)\sum_{u=1}^{N-1}\xi_{t,u}up_t(1-p_t)^{u-1},\\
            \quad\quad\quad\quad\quad\text{if }g\geq\Gamma_t,c = 1,w'=w+1,g'=g,\\
            \sum_{u=1}^{N-1}\xi_{t,u}up_t(1-p_t)^{u-1},\\
            \quad\quad\quad\quad\quad\text{if }0\leq g <\Gamma_t,c= 1,w'\!=\!w+1,g'\!=\!g,\\
            p_t\sum_{u=0}^{N-1}\xi_{t,u}(1-p_t)^{u},\\
            \quad\quad\quad\quad\quad\text{if }g\geq\Gamma_t,c = 1,w'=w+1,g'=0,\\
            \sum_{u=1}^{N-1}\xi_{t,u}\sum_{u'=2}^{u+1}\binom{u+1}{u'}p_t^{u'}(1-p_t)^{u+1-u'},\\
            \quad\quad\quad\quad\quad\text{if }g\geq\Gamma_t,c = *,w' = w+1,g' = g,\\
            \sum_{u=2}^{N-1}\xi_{t,u}\sum_{u'=2}^{u}\binom{u}{u'}p_t^{u'}(1-p_t)^{u-u'},\\
            \quad\quad\quad\quad\quad\text{if }0\leq g <\Gamma_t,c = *,w'\! = \!w+1,g'\! = \!g,\\
          0,\quad\quad\quad\quad\text{otherwise.}
        \end{cases}
    \end{align}
In Eq.~\eqref{eq: Bayesian update1}, the first and second cases correspond to that no devices transmits, 
the third and fourth cases correspond to that one of the other $N-1$ devices transmits successfully,
the fifth case corresponds to that the tagged device transmits successfully when it is active in slot $t$,
the sixth and seventh cases correspond to that a collision occurs. 

\begin{algorithm}[htbp] 
    \caption{The proposed enhanced T-AGDSA.}
    \label{algo1}
    \begin{algorithmic}[1]
    \STATE Set $t=0$. 
        \FOR{each $n\in\mathcal{N}$}
        \STATE // Implemented at the beginning of slot $t$.
             \IF {$t=0$}
                  \STATE  Device $n$ uses Eq.~\eqref{eq: f0} to obtain $\hat{f}_{0,w,g}$.
             \ELSE
                 \IF{$t = mD, m \in \mathbb{Z}^+$}
                     \STATE Device $n$ uses Eq.~\eqref{eq: ft+1} to obtain $(\hat{f}_{t,w,g})_{w,g\in\mathbb{N}}$.
                  \ELSE
                     \STATE Device $n$ obtains \\ $(\hat{f}_{t,w,g})_{w,g\in\mathbb{N}} = (\hat{f}_{(t-1)^+,w,g})_{w,g\in\mathbb{N}}$.
                  \ENDIF
             \ENDIF
             \STATE Based on Eq.~\eqref{eq: pto}, device $n$ chooses $\Gamma_t$ and $p_t$ that maximize Eq.~\eqref{eq: eRt}.
              \IF {$g_{n,t}\geq\Gamma_t$}
                   \STATE In slot $t$, device $n$ transmits with probability $p_t$,
              \ELSE
                   \STATE In slot $t$, device $n$ keeps silent.
              \ENDIF
        \STATE // Implemented at the end of slot $t$.
              \STATE Device $n$ obtains the channel status $c_t$. 
              \STATE Device $n$ uses Eq.~\eqref{eq: Bayesian update} to obtain $(\hat{f}_{t^+,w,g})_{w,g\in\mathbb{N}}$.
        \ENDFOR 
        \STATE $t=t+1$, return to step 2.  
    \end{algorithmic}
\end{algorithm}
\subsection{Computing $(\hat{f}_{t+1,w,g})_{w,g\in\mathbb{N}}$ Using the Update Generation Probability}
It remains to obtain $(\hat{f}_{t+1,w,g})_{w,g\in\mathbb{N}}$ at the beginning of slot $t+1$ based on $(\hat{f}_{t^+,w,g})_{w,g\in\mathbb{N}}$ and the update generation probability $\lambda$.

Initially, for each $n\in\mathcal{N}$, each device knows 
\begin{equation}\label{wn0}
    h_{n,0} = w_{n,0} = g_{n,0} = 0,
\end{equation}
which implies
\begin{align}\label{eq: f0}
    \hat{f}_{0,w,g} = 
    \begin{cases}
    1, &\text{if } w = g = 0,\\
    0, &\text{otherwise.}
    \end{cases}
\end{align}
Considering that each device independently generates an update with probability $\lambda$ at the beginning of each frame $m$ (i.e., $t = mD, m \in \mathbb{Z}^+$) and does not generate updates at other time points, we have

\begin{align}\label{eq: ft+1}
     \hat{f}_{t+1, w', g'} &\triangleq \text{Pr}\big(w_{n,t+1} = w',g_{n,t+1}=g'|\lambda, (\hat{f}_{t^+,w,g})_{w,g\in\mathbb{N}}\big)\notag \\
     &=\sum_{w,g\in\mathbb{N}}\hat{f}_{t^+,w,g}\text{Pr}\Big(w_{n,t+1} = w',g_{n,t+1}=g'|\lambda, \notag \\
    &\quad\quad\quad\quad\quad\quad\quad\quad\quad\quad w_{n,t^+} = w,g_{n,t^+}=g\Big) ,
\end{align}
for each $t,w',g'\in\mathbb{N}$,
where
\begin{align}
   &\text{Pr}\big(w_{n,t+1} = w',g_{n,t+1}=g'|\lambda, w_{n,t^+} = w,g_{n,t^+}=g\big) \notag \\
    &=
    \begin{cases}
        \lambda, &\text{if }t = mD-1, m \in \mathbb{Z}^+, w' = 0, g' = w + g, \\
        1-\lambda,&\text{if }t = mD-1, m \in \mathbb{Z}^+, w' = w, g' = g, \\
        1, &\text{if }t \neq mD-1, m \in \mathbb{Z}^+, w' = w, g' = g, \\
        0, &\text{otherwise.} 
    \end{cases}
\end{align}

The proposed enhanced T-AGDSA is summarized in Algorithm~\ref{algo1}.

\section{Numerical Results}\label{Sec:Numerical Results}
This section consists of three subsections.
The first subsection validates the analytical modeling of the proposed basic T-AGDSA and examines its advantage over the schemes in~\cite{bae2022age,Yavascan2021JSAC,Kadota2021,Yates2017ISIT}.
The second subsection examines the advantage of the proposed enhanced T-AGDSA over the schemes in~\cite{bae2022age,Chen2022TIT,Moradian2024TCOM}.
The third subsection compares the proposed basic T-AGDSA and enhanced T-AGDSA.
The scenarios considered in the simulations are in accordance with the descriptions in Section~\ref{sec:SystemModel}.
We shall vary the network configuration over a wide range to validate our theoretical study.
Each simulation result is obtained from 10 independent simulation runs with $10^7$ slots in each run.

\begin{figure*}[!ht]
  \centering
    \subfloat[$D = 1.$]
      {     
      \label{D1lambda}\includegraphics[width=3.2in]{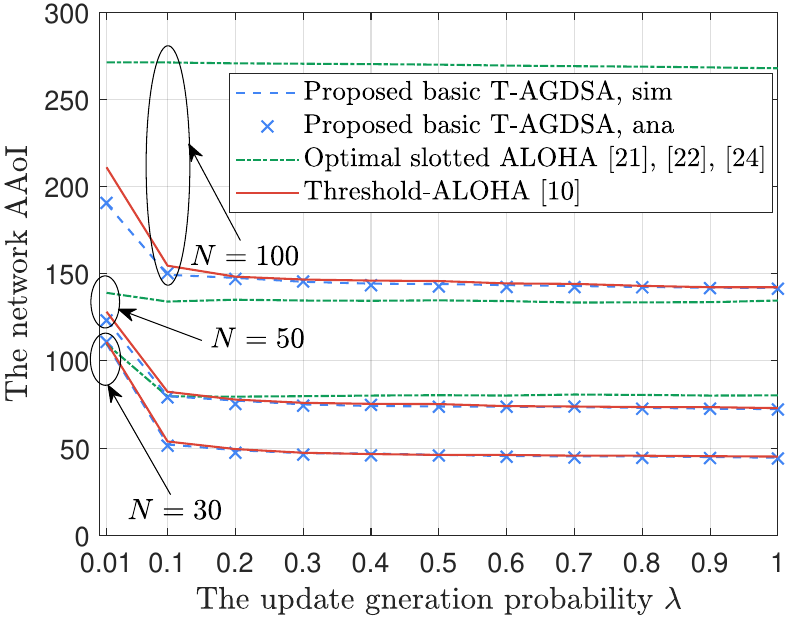}
      }
    \subfloat[$D = 10.$]
      {     
      \label{D10lambda}\includegraphics[width=3.2in]{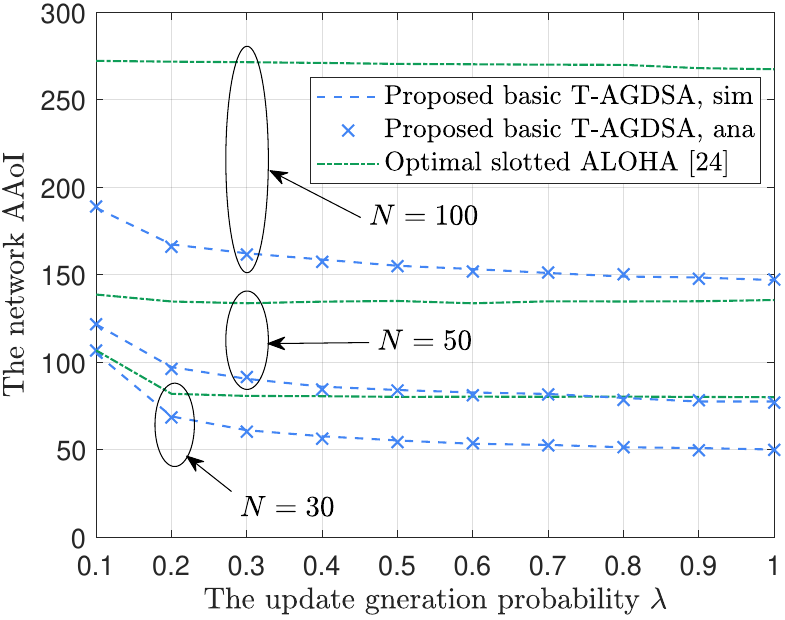}
      }
      \\
     \subfloat[$D = 20.$]
      {
      \label{D20lambda}\includegraphics[width=3.2in]{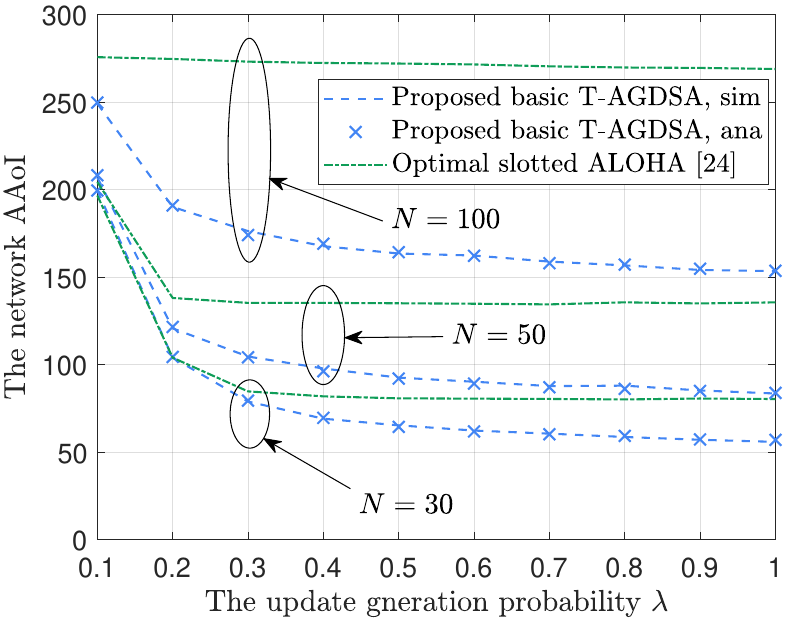}
      }
      \subfloat[$D = 50.$]
      {
      \label{D50lambda}\includegraphics[width=3.2in]{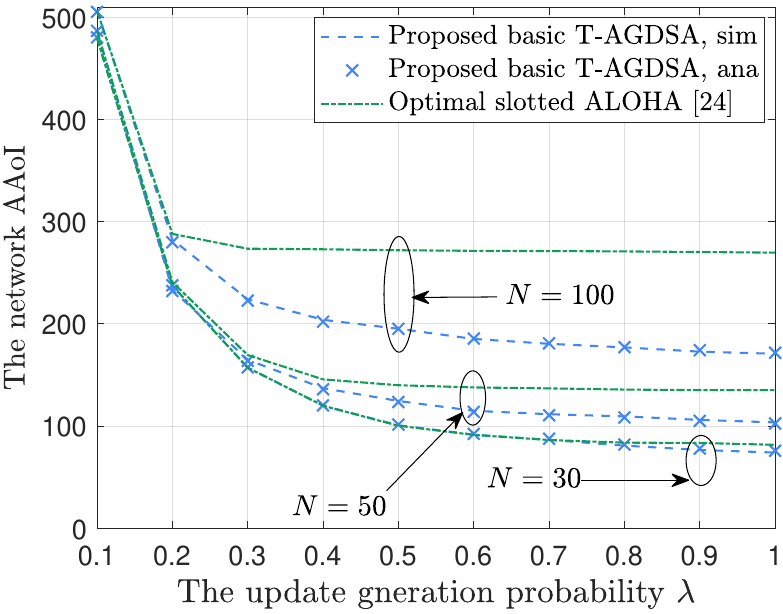}
      }
        \caption{The network AAoI of the proposed basic T-AGDSA versus $\lambda$ for different $N$ and $D$.}
	\label{fig: Dfig}
\end{figure*}

\subsection{Proposed Basic T-AGDSA}
\label{Sec: Numerical Results for the stationary AGDRA}
We consider the following two simply implemented schemes as benchmarks.
\begin{enumerate}
   \item Optimal slotted ALOHA~\cite{bae2022age,Yates2017ISIT,Kadota2021}: under the age gain threshold $\Gamma_t = 1$, each device uses an optimal fixed $p_t$. Note that~\cite{bae2022age} only considered $\lambda=1$ and~\cite{Yates2017ISIT,Kadota2021} only considered $D=1$, so we will obtain the network AAoI for other cases via simulations.  
  \item Threshold-ALOHA for $D=1$~\cite{Yavascan2021JSAC}: each device uses an optimal fixed $\Gamma_t$ and an optimal fixed $p_t$ when $\lambda=D=1$, and uses an suboptimal fixed $\Gamma_t$ and an suboptimal fixed $p_t$ when $\lambda<1$, $D=1$.
\end{enumerate}

Fig.~\ref{fig: Dfig}(a) shows the network AAoI of the proposed basic T-AGDSA as a function of the update generation probability $\lambda$ for different $N$ when $D = 1$.
The curves indicate that our analytical modeling is accurate in all the cases.
We observe that the network AAoI of all the three schemes first decreases with $\lambda$ and then remains almost the same.
This is because larger $\lambda$ is helpful to reduce the AAoI due to the delivery of fresher updates, but this effect would become weaker due to severer contention when $\lambda$ is larger.
We also observe that the proposed basic T-AGDSA enjoys up to $43.59\%$ improvement over the optimal slotted ALOHA~\cite{bae2022age,Yates2017ISIT,Kadota2021}, which verifies the benefit of introducing the fixed threshold $\Gamma^{\text{sta}}$. 
We further observe that the proposed basic T-AGDSA enjoys up to $10.24\%$ improvement over the threshold-ALOHA~\cite{Yavascan2021JSAC} in large-scale networks with sporadic individual traffic (i.e., when $N$ is large and $\lambda$ is small), but performs almost the same in other cases.
This is because the age gain threshold is more helpful in reducing the AAoI compared with the AoI threshold used in the threshold-ALOHA~\cite{Yavascan2021JSAC} and this advantage is notable when $N$ is large and $\lambda$ is small. 
Meanwhile, this advantage is enlarged when $\lambda$ is small because the threshold-ALOHA~\cite{Yavascan2021JSAC} used the assumption of $\lambda=1$ to obtain the transmission policy when $\lambda < 1$.

Fig.~\ref{fig: Dfig}(b)--(d) show the network AAoI of the proposed basic T-AGDSA as a function of $\lambda$ for different $N$ when $D = 10, 20,50$, respectively.
Note that the threshold-ALOHA~\cite{Yavascan2021JSAC} is inapplicable when $D>1$.
The accuracy of our analytical modeling is verified again in these cases.
We observe that, compared with optimal slotted ALOHA~\cite{bae2022age}, the proposed basic T-AGDSA enjoys up to $44.31\%$ improvement when $D = 10$,
up to $41.52\%$ improvement when $D = 20$,
and up to $35.70\%$ improvement when $D = 50$.
These results indicate that introducing $\Gamma^{\text{sta}}$ is effective in improving the AAoI for a wide range of configurations.
We further observe that, in general, the advantage of the proposed basic T-AGDSA diminishes when $\lambda$ decreases, $N$ decreases, or $D$ increases.
This is because the effect of introducing $\Gamma^{\text{sta}}$ to mitigate the contention becomes weaker in these cases.

\subsection{Proposed Enhanced T-AGDSA}
\label{Sec: Numerical Results for the adaptive AGDRA}
We consider the following four ideal or high-computational-overhead schemes as benchmarks.
\begin{enumerate}
\item Ideal scheduling~\cite{Chen2022TIT}: the AP always selects one of the devices with the highest age gains to transmit. Obviously, it provides a lower bound on the network AAoI.
  \item Ideal adaptive slotted ALOHA~\cite{bae2022age}: each device uses $p_t = 1/n_t$ and $\Gamma_t = 1$, where $n_t$ represents the number of active devices in slot $t$.
 \item AAT~\cite{Chen2022TIT} for $D = 1$: see Section~\ref{related work} for details.
 \item Practical T-DFSA~\cite{Moradian2024TCOM} for $D = 1$: see Section~\ref{related work} for details.%
\end{enumerate} 
In addition, we consider another lower bound $D/\lambda+(1-D)/2$ for $D\geq1$ stated in Proposition~1 by assuming no collisions.

\begin{figure}[!ht]
  \centering
  \subfloat[$N = 30.$]
      {     
      \label{N30D1}\includegraphics[width=3.2in]{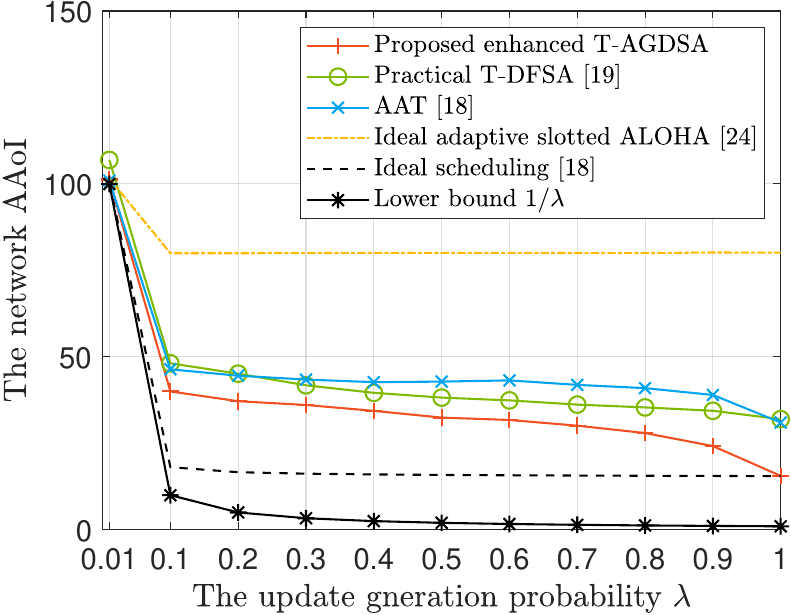}
      }
      \\
    \subfloat[$N = 50.$]
      {     
      \label{N50D1}\includegraphics[width=3.2in]{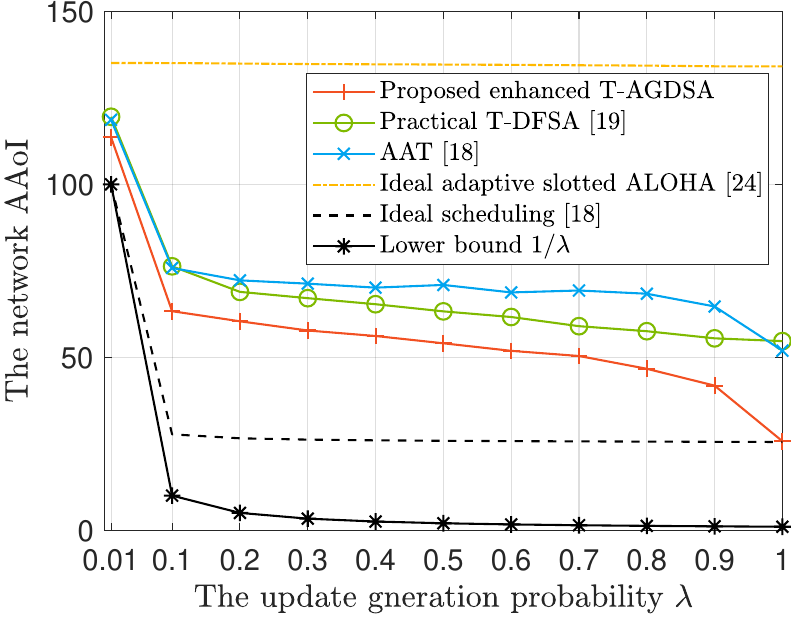}
      }
      \\
      \subfloat[$N = 100.$]
      {
      \label{N100D1}\includegraphics[width=3.2in]{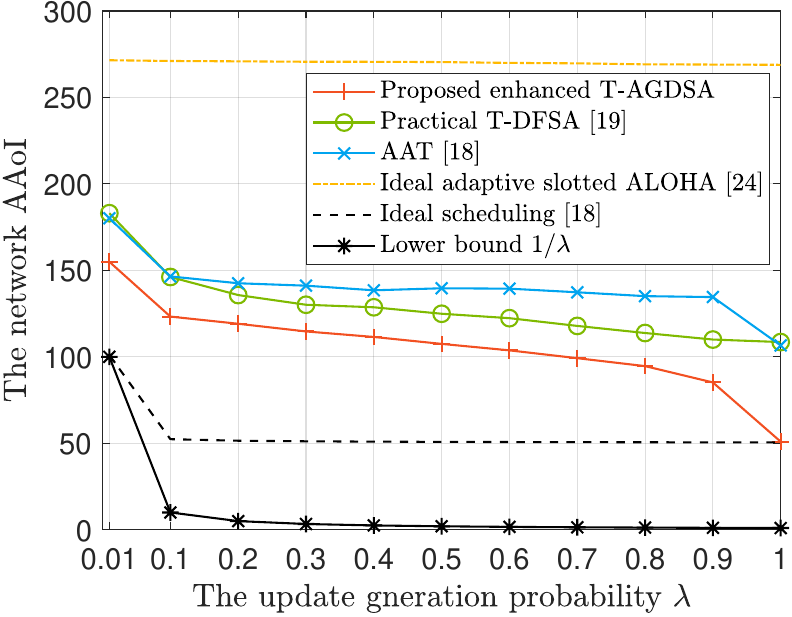}
      }
        \caption{The network AAoI of the proposed enhanced T-AGDSA versus $\lambda$ for different $N$ when $D = 1$.}
	\label{fig: Dfig_A}
\end{figure}

Fig.~\ref{fig: Dfig_A} shows the network AAoI of the proposed enhanced T-AGDSA as a function of $\lambda$ for different $N$ when $D = 1$.
We observe that the network AAoI of the ideal adaptive slotted ALOHA~\cite{bae2022age} first decreases with $\lambda$ and then remains almost the same, while that of the other three schemes always decreases with $\lambda$ due to the effect of introducing the adaptive threshold.
We also observe that the enhanced T-AGDSA enjoys up to $81.06\%$ improvement over the ideal adaptive slotted ALOHA~\cite{bae2022age}.
This verifies the benefit of introducing the adaptive threshold $\Gamma_t$, even if the latter utilizes the ideal knowledge of $n_t$.
We further observe that the enhanced T-AGDSA enjoys up to $52.21\%$ improvement over the AAT~\cite{Chen2022TIT}, and enjoys up to $53.07\%$ improvement over the practical T-DFSA~\cite{Moradian2024TCOM}.
This is owing to our more reasonable $\Gamma_t$, which is computed by not only a more accurate estimation of the age gains (see Remark 6) but also a more reasonable optimization goal (see Remarks 4 and 5).
We also observe that all the schemes enjoy almost the same AAoI (close to $1/\lambda$) when $N\lambda$ is small, which confirms the lower bound $1/\lambda$ proposed in~\cite{Chen2022TIT}.
This is because the inter-arrival time becomes a dominant factor to determine the AAoI when the network traffic is quite low.
On the other hand, when $N\lambda$ is not small, we note that the enhanced T-AGDSA performs closer to the ideal scheduling~\cite{Chen2022TIT} as $\lambda$ increases, which implies that our adaptive threshold can be chosen to limit the contention to fewer devices with higher age gains due to a more accurate estimation of age gains.

\begin{figure}[!ht]
  \centering
    \subfloat[$D = 10.$]
      {     
      \includegraphics[width=3.2in]{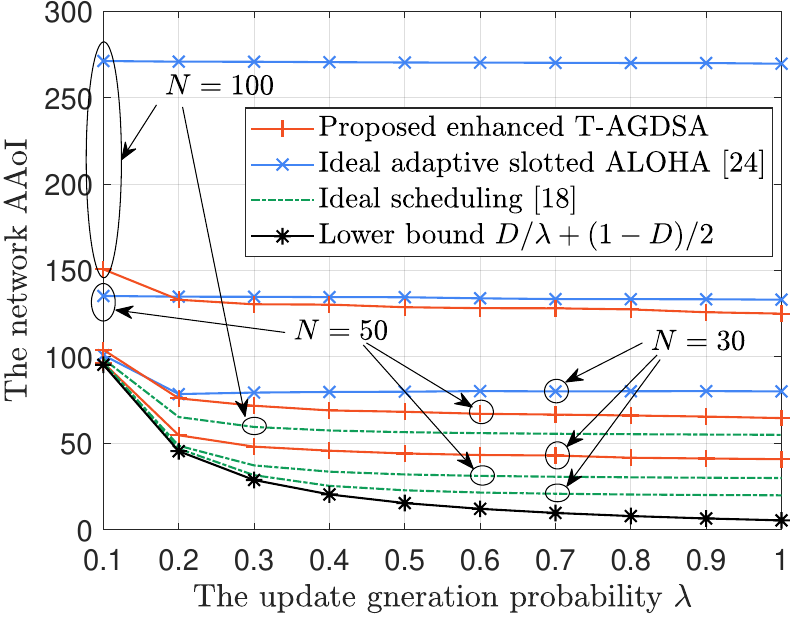}
      }
      \\
    \subfloat[$D = 20.$]
      {     
      \includegraphics[width=3.2in]{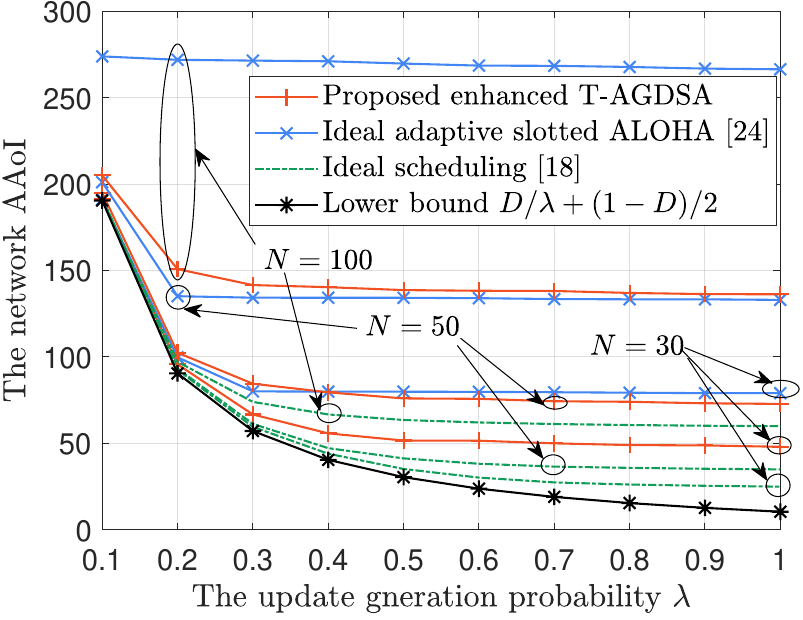}
      }
      \\
     \subfloat[$D = 50.$]
      {
     \includegraphics[width=3.2in]{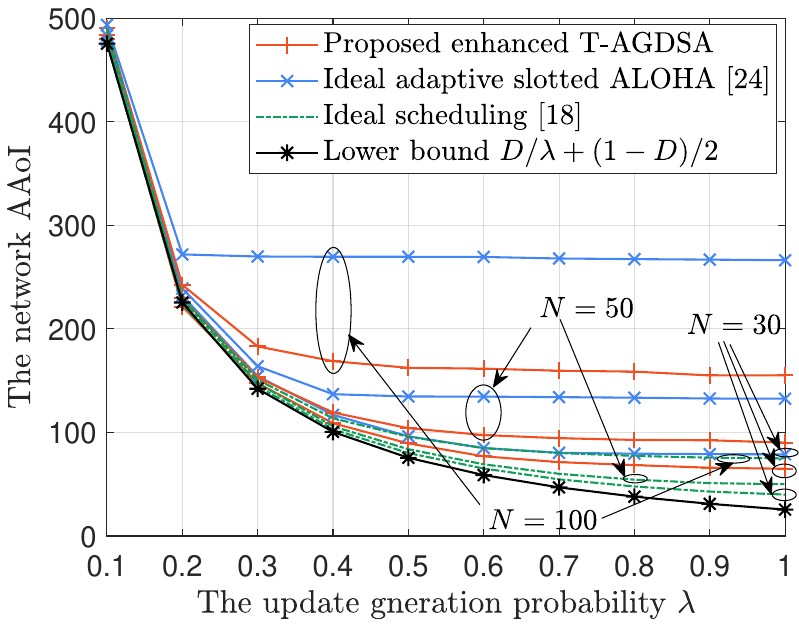}
      }
        \caption{The network AAoI of the proposed enhanced T-AGDSA versus $\lambda$ for different $N$ when $D>1$.}
	\label{fig: Dfig_B}
\end{figure}

Fig.~\ref{fig: Dfig_B} shows the network AAoI of the proposed enhanced T-AGDSA as a function $\lambda$ for different $N$ when $D = 10, 20,50$.
Note that the AAT~\cite{Chen2022TIT} and the practical T-DFSA~\cite{Moradian2024TCOM} are both inapplicable when $D>1$.
We observe that, compared with the ideal adaptive slotted ALOHA~\cite{bae2022age}, the enhanced T-AGDSA enjoys up to $53.61\%$ improvement when $D = 10$, 
up to $48.85\%$ improvement when $D = 20$, 
and up to $41.79\%$ improvement when $D = 50$.
These results indicate that introducing the adaptive $\Gamma_t$ is effective in improving the AAoI for a wide range of configurations.
We further observe that such improvement diminishes when $\lambda$ decreases, $N$ decreases, or $D$ increases.
This is because the effect of introducing the adaptive $\Gamma_t$ to mitigate the contention becomes weaker in these cases.
We also observe that all the schemes enjoy almost the same AAoI (close to $D/\lambda$) when $N\lambda/D$ is small, which confirms our proposed lower bound $D/\lambda+(1-D)/2$.
We also note that the AAoI of the enhanced T-AGDSA always decreases with $\lambda$ when $D = 1$, but first decreases with $\lambda$ and then keeps almost the same when $D > 1$.
This is because, there would be more devices with large age gains as $D$ increases, which leads to severer contentions, thus weakening the advantage of more accurate estimation of age gains under larger $\lambda$.

\subsection{Basic T-AGDSA V.S. Enhanced T-AGDSA}
We observe from Figs.~\ref{fig: Dfig}--\ref{fig: Dfig_B} that, compared with the proposed basic T-AGDSA, the proposed enhanced T-AGDSA enjoys $7.81\%-64.98\%$ improvement when $D = 1$, 
$8.19\%-21.36\%$ improvement when $D = 10$, 
$3.08\%-20.17\%$ improvement when $D = 20$,
and $0.04\%-18.32\%$ improvement when $D = 50$.
As expected, we note that such improvement is close to zero when $N\lambda/D$ is quite low.
We also observe that such improvement always increases with $\lambda$ when $D = 1$, but first increases with $\lambda$ and then decreases with $\lambda$ when $D > 1$.
This indicates again that increased $D$ diminishes the advantage of the enhanced T-AGDSA.
It should be noted that such improvement comes at a cost of higher online computation burden on each device, thus which scheme is preferred depends on the network configurations.

\section{Conclusion}\label{Sec:Conclusion}
In this paper, we have investigated how to design decentralized schemes for reducing the network AAoI in an uplink IoT system with event-driven periodic updating, so that the unavoided contention can be limited to devices with age gains as high as possible.
We proposed a basic T-AGDSA scheme, where the access parameters are fixed and can be obtained offline using the proposed multi-layer Markov modeling approach.
We then proposed an enhanced T-AGDSA scheme, where each device adjusts the access parameters to maximize the estimated network EAR per slot, built on an estimation of the joint probability distribution of local age and age gain of an arbitrary device.
Numerical results validated our theoretical study and confirmed the advantage of our proposed schemes over the existing schemes.
Considering that the enhanced T-AGDSA has higher online computation burden, our work enables one to gain a clear insight into how to choose a suitable T-AGDSA scheme for different network configurations. 
An interesting direction for future research is to design a smarter decentralized scheme under unknown, heterogeneous, and time-varying network configurations.

\appendix
\section*{Proof of Proposition 1}\label{Proof of lower bound}
Suppose that each update can be instantaneously delivered without experiencing collisions.
Let $I_{n,i}$ denote the inter-arrival time between the $(i-1)$-th and $i$-th updates of the tagged device $n$, which is obviously equal to the inter-delivery time.
Considering that each device independently generates an update with probability $\lambda$ at the beginning of each frame and does not generate updates at other time points, we have
\begin{align}\label{eq:Yj}
    \text{Pr}(I_{n,i} = jD) = (1-\lambda)^{j-1}\lambda,
\end{align}
for $i = 1,2,\ldots$, and $j = 1,2,\ldots$.

Since $I_{n,i}/D$ in Eq.~\eqref{eq:Yj} has a geometric distribution with parameter $\lambda$, we have
\begin{align}\label{eq: EY1}
         \mathbb{E}(I_{n,i}) = D\mathbb{E}(I_{n,i}/D) = D/\lambda, 
\end{align}
\begin{align}\label{eq: EY2}
         \mathbb{E}(I^2_{n,i}) = D^2\mathbb{E}\big((I_{n,i}/D)^2\big) = D^2(2-\lambda)/\lambda^2.
\end{align}
Let $\zeta_{n,T}$ be the number of successfully transmitted updates of device $n$ until the $T$-th slot.
The AAoI of device $n$ defined in Eq. \eqref{deqn_ex2a} can be rewritten as
\begin{align}\label{Eq: Dan}
    {\Delta}_n &= \lim_{T\to\infty}\frac{\zeta_{n,T}}{T}\frac{1}{\zeta_{n,T}}\sum_{i=1}^{\zeta_{n,T}}\sum_{h = 1}^{I_{n,i}}h\notag \\
               &= \lim_{T\to\infty}\frac{\zeta_{n,T}}{T}\frac{1}{\zeta_{n,T}}\sum_{i=1}^{\zeta_{n,T}}\big(I_{n,i}(I_{n,i}+1)/2\big)\notag \\
               &=\frac{\mathbb{E}\big(I_{n,i}(I_{n,i}+1)/2\big)}{\mathbb{E}(I_{n,i})}
                =\frac{\mathbb{E}(I^2_{n,i})}{2\mathbb{E}(I_{n,i})} + 1/2.
\end{align}
By substituting Eqs.~\eqref{eq: EY1} and~\eqref{eq: EY2} into Eq.~\eqref{Eq: Dan}, we can obtain  
\begin{align}
   {\Delta}_n = \frac{D^2(2-\lambda)/\lambda^2}{2D/\lambda} + \frac{1}{2} = D/\lambda + (1-D)/2,
\end{align}
which can be served as a lower bound on the network AAoI in our setup.

\bibliography{test}
\end{document}